%% file: deep_pred_lrn_2021.tex
\documentclass[11pt,twoside]{article}
\usepackage[american]{babel}
\usepackage{times,subeqnarray}
\usepackage{url}
\newif\myifpdf
\ifx\pdfoutput\undefined
   \usepackage[dvips]{graphicx}
\else
   \pdfoutput=1        
   \usepackage[pdftex]{graphicx}
\fi
\usepackage{apatitlepages}
\usepackage{setspace} 
\usepackage{psydraft}
\usepackage{amsmath} 
\usepackage[natbibapa]{apacite}   
\newcommand{\oneo}[1]{\frac{1}{#1}}
\usepackage{enumitem}

\myifpdf
  \DeclareGraphicsExtensions{.pdf,.eps,.png,.jpg,.mps,.tif}
\fi

\def\myheading{ Deep Predictive Learning }

\def\mytitle{ Deep Predictive Learning in Neocortex and Pulvinar}

\def\myauthor{Randall C. O'Reilly, Jacob L. Russin, Maryam Zolfaghar, and John Rohrlich\\
  Department of Psychology, Computer Science, and Center for Neuroscience \\
  University of California Davis \\
  1544 Newton Ct\\
  Davis, CA 95618\\
  {\small oreilly@ucdavis.edu}\\}

\def\mynote{
We thank Dean Wyatte, Tom Hazy, Seth Herd, Kai Krueger, Tim Curran, David Sheinberg, Lew Harvey, Jessica Mollick, Will Chapman, Helene Devillez, and the rest of the CCN Lab for many helpful comments and suggestions.
Supported by: ONR grants ONR N00014-19-1-2684 / N00014-18-1-2116, N00014-14-1-0670 / N00014-16-1-2128, N00014-18-C-2067, N00014-13-1-0067, D00014-12-C-0638.  This work utilized the Janus supercomputer, which is supported by the National Science Foundation (award number CNS-0821794) and the University of Colorado Boulder. The Janus supercomputer is a joint effort of the University of Colorado Boulder, the University of Colorado Denver and the National Center for Atmospheric Research.
All data and materials will be available at \url{https://github.com/ccnlab/deep-obj-cat} upon publication.}

\def\myabstract{
How do humans learn from raw sensory experience?  Throughout life, but most obviously in infancy, we learn without explicit instruction.  We propose a detailed biological mechanism for the widely-embraced idea that learning is driven by the differences between predictions and actual outcomes (i.e., \emph{predictive error-driven learning}).  Specifically, numerous weak projections into the pulvinar nucleus of the thalamus generate top-down predictions, and sparse, focal \emph{driver} inputs from lower areas supply the actual outcome, originating in layer 5 intrinsic bursting (5IB) neurons.  Thus, the outcome representation is only briefly activated, roughly every 100 ms (i.e., 10 Hz, \emph{alpha}), resulting in a \emph{temporal difference error signal}, which drives local synaptic changes throughout the neocortex.  This results in a biologically-plausible form of error backpropagation learning.  We implemented these mechanisms in a large-scale model of the visual system, and found that the simulated inferotemporal (IT) pathway learns to systematically categorize 3D objects according to invariant shape properties, based solely on predictive learning from raw visual inputs.  These categories match human judgments on the same stimuli, and are consistent with neural representations in IT cortex in primates.
}

\begin{document}
\bibliographystyle{apacite}


\titlesepage{\mytitle}{\myauthor}{\mynote}{\myabstract}

\pagestyle{myheadings}


The fundamental epistemological conundrum of how knowledge emerges from raw experience has challenged philosophers and scientists for centuries.  Although there have been significant advances in cognitive and computational models of learning \citep{AshbyMaddox11,WatanabeSasaki15,LeCunBengioHinton15} and in our understanding of the detailed biochemical basis of synaptic plasticity \citep{LuscherMalenka12,ShouvalBearCooper02,CooperBear12,UrakuboHondaFroemkeEtAl08}, there is still no widely-accepted answer to this puzzle that is clearly supported by known biological mechanisms and also produces effective learning at the computational and cognitive levels.  The idea that we learn via an active \emph{predictive} process was advanced by Helmholtz in his \emph{recognition by synthesis} proposal \citep{Helmholtz67}, and has been widely embraced in a range of different frameworks \citep{Elman90,ElmanBatesKarmiloff-SmithEtAl96,Mumford92,KawatoHayakawaInui93,DayanHintonNealEtAl95,RaoBallard99,Friston05,HawkinsBlakeslee04,GeorgeHawkins09,Clark13,SummerfielddeLange14,deLangeHeilbronKok18}.  

Here, we propose a detailed biological mechanism for a specific form of \emph{predictive error-driven learning} based on distinctive patterns of connectivity between the neocortex and the higher-order nuclei of the thalamus (i.e., the pulvinar) \citep{ShermanGuillery06,UsreySherman18}.  We hypothesize that learning is driven by the difference between top-down predictions, generated by numerous weak projections into the thalamic relay cells (TRCs) in the pulvinar, and the actual outcomes supplied by sparse, strong \emph{driver} inputs from lower areas.  Because these driver inputs originate in layer 5 intrinsic bursting (5IB) neurons, the outcome is only briefly activated, roughly every 100 ms (i.e., 10 Hz, \emph{alpha}).  Thus, the prediction error is a \emph{temporal difference} in activation states over the pulvinar, from an earlier prediction to a subsequent burst of outcome.  This temporal difference can drive local synaptic changes throughout the neocortex, supporting a biologically-plausible form of error backpropagation that improves the predictions over time \citep{OReilly96,AckleyHintonSejnowski85,HintonMcClelland88,BengioMesnardFischerEtAl17,WhittingtonBogacz19,LillicrapSantoroMarrisEtAl20}.
The temporal-difference form of error-driven learning contrasts with prevalent alternative hypotheses that require a separate population of neurons to compute a prediction error \emph{explicitly} and transmit it directly through neural firing \citep{RaoBallard99,KawatoHayakawaInui93,Friston05,Friston10,OudenKokLange12,LotterKreimanCox16}.

In the following, our primary objective is to describe the hypothesized biologically based mechanism for predictive error-driven learning, contrast it with other existing proposals regarding the functions of this thalamocortical circuitry and other ways that the brain might support predictive learning, and evaluate it relative to a wide range of existing anatomical and electrophysiological data.  We provide a number of specific empirical predictions that follow from this functional view of the thalamocortical circuit, which could potentially be tested by current neuroscientific methods.  Thus, this work proposes a clear functional interpretation of this distinctive thalamocortical circuitry that contrasts with existing ideas in testable ways.

A second major objective is to implement this predictive error-driven learning mechanism in a large-scale computational model that faithfully captures its essential biological features, to test whether the proposed learning mechanism can drive the formation of cognitively useful representations.  In particular, we ask a critical question for any predictive-learning model: can it develop high-level, abstract representations while learning from nothing but predicting low-level visual inputs.  Most visual object recognition models that provide a reasonable fit to neurophysiological data rely on large human-labeled datasets to explicitly train abstract category information via error-backpropagation \citep{CadieuHongYaminsEtAl14,Khaligh-RazaviKriegeskorte14,RajalinghamIssaBashivanEtAl18}.  Thus, it is perhaps not too surprising that the higher layers of these models, which are closer to these category output labels, exhibited a greater degree of categorical organization.

Through large-scale simulations based on the known structure of the visual system, we found that our biologically based predictive learning mechanism developed high-level, abstract representations that significantly diverge from the similarity structure present in the lower layers of the network, and systematically categorize 3D objects according to invariant shape properties.  Furthermore, we found in an experiment using the same stimuli that these categories match human similarity judgments, and that they are also qualitatively consistent with neural representations in inferotemporal (IT) cortex in primates \citep{CadieuHongYaminsEtAl14}.  In addition, we show that comparison predictive backpropagation models lacking these biological features \citep{LotterKreimanCox16} did not learn object categories that go beyond the visual input structure.  Thus, there may be some important features of the biologically based model that enable this ability to learn higher-level structure beyond that of the raw inputs.

It is important to emphasize that our objectives for these simulations are \emph{not} to produce a better machine-learning (ML) algorithm \emph{per se}, but rather to test whether our biologically based model can capture some of the known high-level, cognitive phenomena that the mammalian brain learns.  Thus, we explicitly dissuade readers from the inevitable desire to evaluate the importance of our model based on differences in narrow, performance-based ML metrics.  As discussed later, there are various engineering-level issues regarding the biologically based model's computational cost and performance, that currently limit its ability to compete with simpler, much larger-scale backpropagation models, but we do not think these are relevant to the evaluation of the scientific questions of relevance here.  In short, this model is an instantiation of a scientific theory, and it should be evaluated on its ability to explain a wide range of data across multiple levels of analysis, just as every other scientific theory is evaluated.

The remainder of the paper is organized as follows.  First, we provide a concise overview of the biologically based predictive error-driven learning framework, including the most relevant neural data.  Then, we present a small-scale implementation of the model that learns a probabilistic grammar, to illustrate the basic computational mechanisms of the theory.  This is followed by the large-scale model of the visual system, which learns by predicting over brief movies of 3D objects rotating and translating in space.  We evaluate this model and compare it to two other predictive learning models that directly use error-backpropagation, based on current deep convolutional neural network (DCNN) mechanisms.  Then, we circle back to discuss the relevant biological data in greater detail, along with testable predictions that can differentiate this account from other existing ideas.  Finally, we conclude with a discussion of related models and outstanding issues.

\section{Predictive Error-driven Learning in the Neocortex and Pulvinar}

\begin{figure}
  \centering\includegraphics[width=4in]{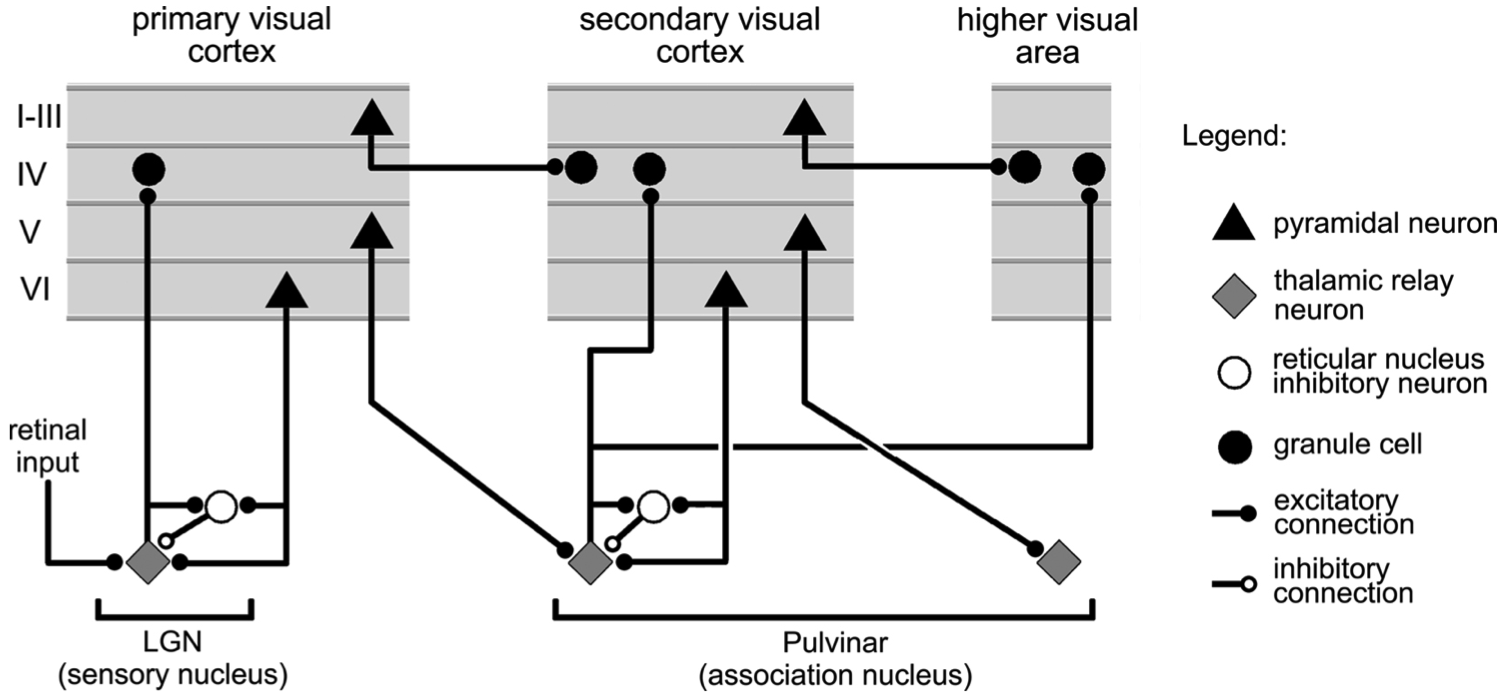}
  \caption{\footnotesize Summary figure from Sherman \& Guillery (2006) showing the strong feedforward \emph{driver} projection emanating from layer 5IB cells in lower layers (e.g., V1), and the much more numerous feedback ``modulatory'' projection from layer 6CT (corticothalamic) cells.  We interpret these same connections as providing a prediction (6CT) vs. outcome (5IB) activity pattern over the pulvinar.}
  \label{fig.sg06}
\end{figure}

\begin{figure}
  \centering\includegraphics[width=6in]{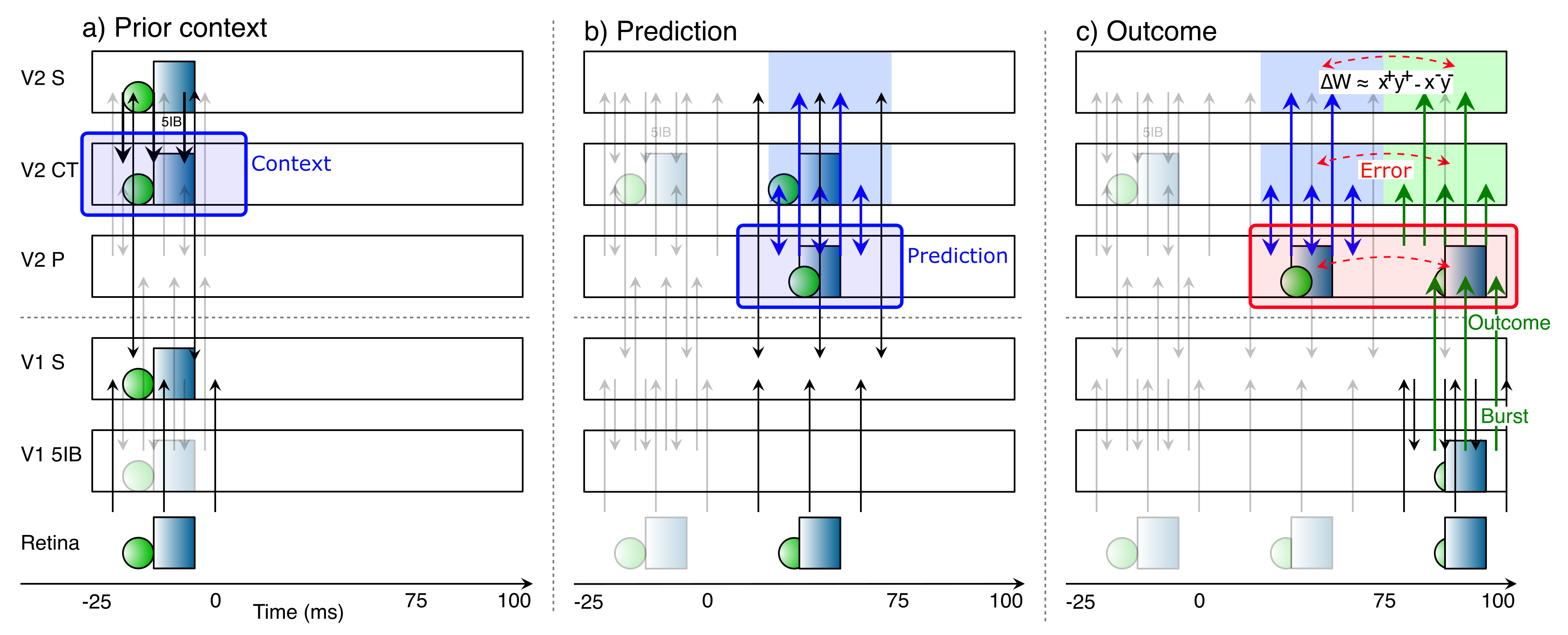}
  \caption{\footnotesize Corticothalamic information flow under our predictive learning hypothesis, shown as a sequence of movie frames (Retina), illustrating the three key steps taking place within a single 125 ms time window, broken out separately across the three panels: {\bf a)} prior context is updated in the V2 CT layer; {\bf b)} which is then used to generate a prediction over the pulvinar (V2 P); {\bf c)} against which the outcome, driven by bottom-up 5IB bursting, represents the prediction error \emph{as a temporal difference between the prediction and outcome states over the pulvinar}. Changes in synaptic weights (learning) in all superficial (S) and CT layers are driven from the \emph{local} temporal difference experienced by each neuron, using a form of the contrastive hebbian learning (CHL) term as shown, where the `$+$' superscripts indicate outcome activations, and `$-$' superscripts indicate prediction. CHL approximates the backpropagated prediction error gradient experienced by each neuron (O'Reilly, 1996), reflecting both direct pulvinar error signals, and indirect corticocortical error signals as well.  In specific: {\bf a)} CT context updating occurs via 5IB bursting (not shown) in higher layer (V2) during prior alpha (100 ms) cycle --- this context is maintained in the CT layer and used to generate predictions. {\bf b)} The prediction over pulvinar is generated via numerous top-down CT projections. This prediction state also projects up to S and CT layers, and from S to all other S layers via extensive bidirectional connectivity, so their activation state reflects this prediction as well.  {\bf c)} The subsequent outcome drives pulvinar activity bottom-up via V1 5IB bursting, and is likewise projected to S and CT layers, ensuring that the relevant temporal difference error signal is available locally in cortex. The difference in activation values across these two time points, in S and CT layers throughout the network, drives learning to reduce prediction errors.  Note that the single most important property of the 5IB bursting is that these driver cells are \emph{not} active during the prediction phase --- the bursting itself may also be useful in the driving property, but that is a secondary consideration to the critical feature of having a time when the prediction alone can be projected onto the pulvinar.}
  \label{fig.dltime}
\end{figure}

Figure~\ref{fig.sg06} shows the thalamocortical circuits characterized by \citet{ShermanGuillery06} (see also \citealp{ShermanGuillery13,UsreySherman18}), which have two distinct projections converging on the principal thalamic relay cells (TRCs) of the \emph{pulvinar}, the primary thalamic nucleus that is interconnected with higher-level posterior cortical visual areas \citep{Shipp03,ArcaroPinskKastner15,HalassaKastner17}.  One projection consists of numerous, weaker connections originating in deep layer VI of the neocortex (the 6CT corticothalamic projecting cells), which we hypothesize generate a top-down prediction on the pulvinar.  The other is a sparse, focal \citep{Rockland98a,Rockland96} and strong \emph{driver} pathway that originates from lower-level layer 5 intrinsic bursting cells (5IB), which we hypothesize provide the outcome.  These 5IB neurons fire discrete bursts with intrinsic dynamics having a period of roughly 100 ms between bursts \citep{ConnorsGutnickPrince82,SilvaAmitaiConnors91,LarkumZhuSakmann99,FranceschettiGuatteoPanzicaEtAl95,SaalmannPinskWangEtAl12}, which is thought to drive the widely-studied \emph{alpha} frequency of $\sim$ 10 Hz that originates in cortical deep layers and has important effects on a wide range of perceptual and attentional tasks \citep{BuffaloFriesLandmanEtAl11,VanRullenKoch03,MathewsonGrattonFabianiEtAl09,JensenBonnefondVanRullen12,ClaytonYeungKadosh18}.  Critically, unlike many other such bursting phenomena, this 5IB bursting occurs in awake animals \citep{LuczakBarthoHarris09,LuczakBarthoHarris13,SakataHarris09,SakataHarris12}, consistent with presence of alpha in awake behaving states.

The existing literature generally characterizes the 6CT projection as \emph{modulatory} \citep{ShermanGuillery13,UsreySherman18}, but a number of electrophysiological recordings from awake, behaving animals clearly show sustained, continuous patterns of neural firing in pulvinar TRC neurons, which is not consistent with the idea that they are only being driven by their phasic bursting 5IB inputs \citep{Bender82,PetersenRobinsonKeys85,BenderYouakim01,Robinson93,SaalmannPinskWangEtAl12,KomuraNikkuniHirashimaEtAl13,ZhouSchaferDesimone16}.  Indeed, these recordings show that pulvinar neural firing generally resembles that of the visual areas with which they interconnect, in terms of neural receptive field properties, tuning curves, etc.  This is important because our predictive learning framework requires that these 6CT top-down projections be capable of directly driving TRC activity.  Specifically, in contrast to the standard view, the core idea behind our theory is that the top-down 6CT projections drive a \emph{predicted} activity pattern across the extent of the pulvinar, which precedes the subsequent \emph{outcome} activation state driven by the strong 5IB inputs.

Figure~\ref{fig.dltime} illustrates the temporal evolution of activity states according to our predictive learning theory, which is somewhat challenging to convey because the critical signals driving learning unfold \emph{over time} \citep{KachergisWyatteOReillyEtAl14,OReillyWyatteRohrlich14,OReillyWyatteRohrlich17}.  We hypothesize that synaptic plasticity throughout the cortex is sensitive to the resulting \emph{temporal differences} that emerge initially in the pulvinar.  Thus, unlike other models (as we discuss in depth later) the prediction error here is not captured directly in the firing of a special population of error-coding neurons, but rather remains as a temporal difference error signal.

The figure shows a single 125 ms time window of a 100 ms alpha cycle for the purposes of illustration (the actual timing is likely to be more dynamic as discussed next).  The activity state in pulvinar TRC neurons, representing a prediction, as driven by the top-down 6CT projections, should develop during the first $\sim$ 75 ms, when the 5IB neurons are paused between bursting.  Then the final $\sim$ 25 ms largely reflects the strong 5IB bottom-up ground-truth driver inputs when they burst.  Thus, the prediction error signal is reflected in the temporal difference of these activation states as they develop over time.  In other words, our hypothesis is that the pulvinar is directly representing either the top-down prediction or the bottom-up outcome at any given time, and the temporal difference between these states implicitly encodes a prediction error.  While the deep 6CT layer is involved in generating a top-down prediction over the pulvinar, the superficial layer neurons continuously represent the current state, simultaneously incorporating bottom-up and top-down constraints via their own connections with other areas.  To ensure that the prediction is not directly influenced by this current state representation (i.e., ``peeking at the right answer''), it is important that the 6CT neurons encode temporally delayed information, consistent with available data \citep{HarrisShepherd15,SakataHarris09,Thomson10}.

The actual biological system is likely to be much more dynamic than the simplistic cartoon with rigid 100 ms timing as shown in Figure~\ref{fig.dltime}, based on a set of neural mechanisms that can work together to enable it to more flexibly entrain the predictive learning cycle to the environment.  These mechanisms would also tend to increase activity and learning associated with unexpected outcomes relative to expected ones, consistent with the observed \emph{expectation suppression} phenomena \citep{SummerfieldTrittschuhMontiEtAl08,TodorovicEdeMarisEtAl11,MeyerOlson11,BastosUsreyAdamsEtAl12}.

Specifically, various underlying mechanisms result in neural \emph{adaptation}, which is generally thought to increase neural activity and learning associated with novel inputs relative to recently familiar ones \citep{MullerMethaKrauskopfEtAl99,AbbottVarelaSenEtAl97,BretteGerstner05,Grill-SpectorHensonMartin06,Hennig13}.  In the case where outcomes are consistent with prior predictions (i.e., the predictions are accurate), the same population of neurons across pulvinar and cortex should be active over time, whereas unpredicted outcomes will generally activate new subsets of neurons in superficial cortical layers representing the current state.  Thus, due to adaptation, there should be a phasic increase in activity in these superficial neurons at the onset of unpredicted stimuli relative to predicted ones.  Furthermore, the 5IB neurons downstream of these superficial neurons may be particularly responsive to these phasic activity increases, causing their bursting to coincide preferentially with unexpected outcomes, thereby driving the phase resetting of the alpha cycle to such events. Thus, during a sequence of predicted states, the pulvinar may experience relatively weaker or even absent 5IB driving inputs, until an unpredicted stimulus arises.  At this point, error-driven learning would be more strongly engaged as a function of the phasic release from adaptation and 5IB burst activation.  We discuss these dynamics more later in the context of the comparison with explicit error coding models.

We also hypothesize that 5IB bursting preferentially drives the synaptic plasticity processes to take place at that time, due the strong driving nature of the outputs from these neurons.  In computational terms originating with the \emph{Boltzmann Machine} \citep{AckleyHintonSejnowski85,HintonSalakhutdinov06}, this anchors the target or \emph{plus} phase to be at this point of 5IB bursting.  Furthermore, this means that the predictive nature of the prior minus phase naturally emerges just by virtue of it being the state prior to 5IB bursting: the learning rule automatically causes that prior state to better anticipate the subsequent state.  Thus, even if no prediction was initially generated, learning over multiple iterations will work to create one, to the extent that a reliable prediction can be generated based on internal states and environmental inputs.  Likewise, assuming relevant activity traces naturally persist over timescales longer than the alpha cycle, this predictive learning process can take advantage of any such remaining traces to learn across these longer timescales, even though it is operating at the faster alpha scale.

In short, learning always happens whenever something unexpected occurs, at any point, and drives the development of predictions immediately prior, to the extent such predictions are possible to generate.  In the typical lab experiment where phasic stimuli are presented without any predictable temporal sequence (which is uncharacteristic of the natural world), there may often be no significant prediction prior to stimulus onset, and we would expect such stimuli to reliably drive 5IB bursting, which is consistent with available electrophysiological data \citep{Bender82,PetersenRobinsonKeys85,BenderYouakim01,Robinson93,LuczakBarthoHarris09,LuczakBarthoHarris13,KomuraNikkuniHirashimaEtAl13,ZhouSchaferDesimone16}.  Thus, unlike Figure~\ref{fig.dltime}, such situations would \emph{start} with a 5IB-triggered plus phase, without a significant minus phase prior to that.

As may be evident by this point, we are mainly focused on \emph{prediction} in the sense of the humorous quote: ``prediction is very difficult, especially about the future'' (attributable to Danish author Robert Storm Petersen), whereas this term is potentially confusingly used in a much broader sense in most Bayesian-inspired predictive coding frameworks \citep{RaoBallard99,Friston05,deLangeHeilbronKok18}.
These frameworks use ``prediction'' to encompass everything from genetic biases to the results of learning in the feedforward synaptic pathways to top-down filling-in or biasing of the current stimulus properties, and fairly rarely use it in the ``about the future'' sense.   We think these different phenomena are each associated with different neural mechanisms at different time scales \citep{OReillyHazyHerd16,OReillyWyatteHerdEtAl13,OReillyMunakataFrankEtAl12}, and thus prefer to treat them separately, while also recognizing that they can clearly interact as well.

Thus, our use of the term \emph{prediction} here refers specifically to \emph{anticipatory} neural firing that predicts subsequent stimuli.  We use the term \emph{postdiction} to refer to the operation of this predictive mechanism after a stimulus has been initially processed (to consolidate and more deeply encode, as in an auto-encoder model), and distinguish both from \emph{top-down excitatory biasing}, which directly influences the online superficial layer neural representations of the current stimulus \citep{DesimoneDuncan95,ReynoldsChelazziDesimone99,MillerCohen01,OReillyWyatteHerdEtAl13}.  Finally, many discussions of prediction error in the literature include late, frontally-associated processes such as those associated with the P300 ERP component \citep{HolroydColes02}.  We specifically exclude these from the scope of the mechanisms described here, which are anticipatory, fast, and low-level, as is appropriate for the posterior cortical sensory processing areas that interconnect with the pulvinar.

\subsection{Computational Properties of Predictive Learning in the Thalamocortical Circuits}

We next elaborate the connections between the computational properties required for predictive learning, and the properties of the circuits interconnecting cortex and the pulvinar, which appear to be notably well suited for their hypothesized role in predictive learning.  We begin with a relatively established interpretation of superficial layer processing, to contextualize subsequent points about the special functions required of the deep layers and the thalamus.

\begin{itemize}

	\item {\bf The superficial cortical layers continuously represent the current state:}  The superficial layer pyramidal neurons are densely and bidirectionally interconnected with other cortical areas, and update quickly to new stimulus inputs, with continuous, relatively rapid firing (i.e., up to about 100 Hz for preferred stimuli).  These neurons integrate higher-level top-down information with bottom-up sensory information to resolve ambiguities, focus attention, fill in missing information, and generally enhance the consistency and quality of the online representations \citep{RumelhartMcClelland82,Hopfield84,DesimoneDuncan95,ReynoldsChelazziDesimone99,MillerCohen01,OReillyWyatteHerdEtAl13,OReillyMunakataFrankEtAl12,OReillyHazyHerd16}. As noted above, we distinguish this form of top-down processing, which is often most evident during the period \emph{after} stimulus onset \citep{LeeMumford03}, from the specifically predictive, anticipatory sort.

	\item {\bf Predictions must be insulated against receiving current state information (it isn't prediction if you already know what happens):} Given that the superficial layers are continuously updating and representing the current state, some kind of separate neural system insulated from this current state information must be used to generate predictions, otherwise the prediction system can just ``cheat'' and directly report the current state.  It may seem counter-intuitive, but making the prediction task \emph{harder} is actually beneficial, because that pushes the learning to capture deeper, more systematic regularities about how the environment evolves over time.  In other words, like any kind of cheating, the cheater itself is cheated because of the reduced pressure to learn, and learning is the real goal.  

	\item {\bf Predictions take time and space to generate:} Non-trivial predictions likely require the integration of multiple converging inputs from a range of higher-level cortical areas, each encoding different dimensions of relevance (e.g., location, motion, color, texture, shape, etc).  Thus, sufficient time and space (i.e., neural substrates with relevant connectivity) must be available to integrate these signals into a coherent predicted state, and per the above point, these substrates must be separated from the influence of current state information.  This fits with the properties of the layer 6CT neurons and their deep layer inputs, which we hypothesize are insulated from superficial-layer firing by virtue of being driven locally by the 5IB bursting within their own cortical microcolumn, such that the inter-bursting pause period provides a time window when these deep layers can integrate and generate the prediction.
	
	Biologically, this is consistent with the delayed responses of 6CT neurons \citep{HarrisShepherd15,SakataHarris09,Thomson10}.  Computationally, these neurons function much like the simple recurrent network (SRN) context layer updating \citep{Elman90,Jordan89} which reflects the prior trial's state, as discussed in detail in the Appendix.  The overall duration of the alpha cycle may represent a reasonable compromise between the prediction integration time and the need to keep up with predictions tracking changes in the world.  Notably, films are typically shown at just over 2 times the alpha frequency (24 Hz), suggesting a Nyquist sampling relative to the underlying alpha processing.

	\item {\bf The predicted state must be directly aligned with the outcome state it predicts:} A prediction error is a difference between two states, so these prediction and outcome states must be directly comparable such that their difference meaningfully represents the actual prediction error, and not some other kind of irrelevant encoding differences.  In other words, the prediction and the outcome must be represented in the same ``language'', so that the ``words'' from the prediction can be directly compared against those of the outcome --- if the prediction was in Japanese and the outcome in English, it would be hard to tell whether the prediction was correct or not!  Thus, a common neural substrate with two different input pathways is required, one reflecting the prediction and the other the outcome, so that both converge onto the same representational system within this common neural substrate.  This fits well with the two pathways converging into the pulvinar: the 6CT top-down prediction-generation pathway and the lower-level 5IB driving inputs.

	\item {\bf The outcome signal should be as \emph{veridical} as possible (i.e., directly reflecting the bottom-up outcome), and should arise from lower areas in the hierarchy relative to the corresponding predictive 6CT inputs:}  Given that the outcome is the driver of learning, if it were to be corrupted or inaccurate, then everything that is learned would then be suspect.  To the extent that delusional thinking is present in all people (some moreso than others perhaps) this principle must be violated at some level, but for the lowest levels of the perceptual system at least, it is important that strongly grounded, accurate training signals drive learning.  The bottom-up, sparse, strongly driving nature of the 5IB projections to the pulvinar can directly convey such veridical outcome signals, and ensure that they dominate the activation of their TRC targets.  Based on indirect available data, it is likely that each pulvinar TRC neuron receives only roughly 1-6 driver inputs \citep{ShermanGuillery06,ShermanGuillery11}, such that these sparse inputs directly convey the signal from lower layers, without much further mixing or integration (which could distort the nature of the signal).  Furthermore, these inputs are likely not plastic \citep{UsreySherman18}, again consistent with a need for unaltered, veridical signals.  Lastly, the TRC neurons are distinctive in having no significant lateral interconnectivity \citep{ShermanGuillery06}, enabling them to faithfully represent their inputs.  These properties led \citet{Mumford91} to characterize the pulvinar as a \emph{blackboard}, and we further suggest the metaphor of a \emph{projection screen} upon which the predictions are projected.

	\item {\bf The prediction error must drive learning to reduce subsequent prediction errors:}  Obviously, this is the goal of prediction error learning in the first place, and given that the cortex is what generates predictions, it must be capable of learning based on prediction error signals represented over the pulvinar.  Computationally, the critical problem here is \emph{credit assignment}: how do the error signals direct learning in the proper direction for each individual neuron, to reduce the overall prediction error?  The error backpropagation procedure solves this problem \citep{RumelhartHintonWilliams86}, but requires biologically implausible retrograde signaling across the entire network of neural communication \citep{Crick89}, to propagate the error proportionally back along the same channels that drive forward activation.  Bidirectional connections, which are ubiquitous in the cortex \citep{MarkovErcsey-RavaszGomesEtAl14,FellemanVanEssen91} and computationally beneficial for other reasons as noted earlier, can eliminate that problem by ``implicitly'' propagating error signals via standard neural communication mechanisms along both directions of connectivity \citep{OReilly96}.
	
	This solution to the credit assignment problem relies on a \emph{temporal difference} error signal, as originally developed for the \emph{Boltzmann machine} \citep{AckleyHintonSejnowski85}.  The bidirectional neural communication at one point in time is encoding and sharing the prediction among the entire network of neurons.  Then, this same network of connections is reused at another point in time to encode and communicate the outcome.  Mathematically, the difference in activation state across these two points in time, locally at each individual neuron, provides an accurate estimate of the error backpropagation gradient \citep{OReilly96}.  In effect, this temporal difference tells each neuron which direction it needs to change its activation state to reduce the overall error.  The reuse of the very same network of connections across both points in time ensures the overall alignment of the two activation states, as noted above, such that this temporal difference precisely represents the error signal.  While various other schemes for error-driven learning in biologically-plausible networks have been proposed \citep[e.g.,][]{BengioMesnardFischerEtAl17,WhittingtonBogacz19,LillicrapSantoroMarrisEtAl20}, the temporal-difference framework with bidirectional connectivity provides a particularly good fit with the natural temporal ordering of predictive learning (prediction then outcome) and the extensive bidirectional connectivity of the thalamocortical circuits \citep{Shipp03}.
	
	\item {\bf Temporal differences in activation state across the alpha cycle, between prediction and outcome states, must drive synaptic plasticity:}  The final step needed to connect all of the elements above is that neurons actually modify their synaptic strengths in proportion to the temporal-difference error signal.  We have recently provided a fully explicit mechanism for this form of learning \citep{OReillyMunakataFrankEtAl12}, based on a biologically-detailed model of spike timing dependent plasticity (STDP) \citep{UrakuboHondaFroemkeEtAl08}.  We showed that when activated by realistic Poisson spike trains, this STDP model produces a nonmonotonic learning curve similar to that of the BCM model \citep{BienenstockCooperMunro82}, which results from competing calcium-driven postsynaptic plasticity pathways \citep{ShouvalBearCooper02,CooperBear12}.  As in the BCM framework, we hypothesized that the threshold crossover point in this nonmonotonic curve moves dynamically --- if this happens on the alpha timescale \citep{LimMcKeeWoloszynEtAl15}, then it can reflect the prediction phase of activity, producing a net error-driven learning rule based on a subsequent calcium signal reflecting the outcome state.  The resulting learning mechanism naturally supports a combination of both BCM-style hebbian learning and error-driven learning, where the BCM component acts as a kind of regularizer or bias, similar to weight decay \citep{OReillyMunakata00,OReillyMunakataFrankEtAl12}.
\end{itemize}

Thus, remarkably, the pulvinar and associated thalamocortical circuitry appears to provide \emph{precisely} the necessary ingredients to support predictive error-driven learning, according to the above analysis.  Interestingly, although \citet{ShermanGuillery06} did not propose a predictive learning mechanism as just described, they did speculate about a potential role for this circuit in motor forward-model learning and the predictive remapping phenomenon \citep{ShermanGuillery11,UsreySherman18}.  In addition, \citet{PennartzDoraMuckliEtAl19} also suggested that the pulvinar may be involved in predictive learning, but within the explicit error-coding framework and not involving the detailed aspects of the above-described circuitry.

It bears emphasizing the synergy between the various considerations above for the benefits of the pause in 5IB firing between bursts.  First, this pause is critical for creating the time window when the predictive network is representing and communicating the prediction state, without influence from the outcome state.  Further, it creates the temporal difference in activation state in the pulvinar between prediction and outcome, which is needed for driving error-driven learning.  Thus, for both the 6CT and pulvinar layers, the periodic pausing of 5IB neurons is essential for creating the predictive learning dynamic.  Interestingly, by these principles, the lack of such burst / pause dynamics in the driver inputs to first-order sensory thalamus areas such as the LGN and MGN \citep{ShermanGuillery06} means that these areas should \emph{not} be directly capable of error-driven predictive learning.  This is consistent with a number of models and theoretical proposals suggesting that primary sensory areas may learn predominantly through hebbian-style self-organizing mechanisms \citep{Miller94,Bednar12}.  Nevertheless,  primary sensory areas do receive ``collateral'' error signals from the pulvinar \citep{Shipp03}, which could provide some useful indirect error-driven learning signals.

Note that this form of temporal-difference learning signal is distinct from the widely-used TD (temporal-difference) model in reinforcement learning \citep{SuttonBarto98}, which is scalar, and applies to reward expectations, not sensory predictions (although see \citealp{GardnerSchoenbaumGershman18} and \citealp{Dayan93} for potential connections between these two forms of prediction error).  Finally, as we discuss later, this proposed predictive role for the pulvinar is compatible with the more widely-discussed role it may play in attention \citep{LaBergeBuchsbaum90,BenderYouakim01,SnowAllenRafalEtAl09,SaalmannKastner11,ZhouSchaferDesimone16,FiebelkornKastner19}.  Indeed, we think these two functions are synergistic (i.e., you predict what you attend, and vice-versa; \citealp{RichterdeLange19}), and have initial computational results consistent with this idea.

\section{Predictive Learning of Temporal Structure in a Probabilistic Grammar}

\begin{figure}
  \centering\includegraphics[width=2.5in]{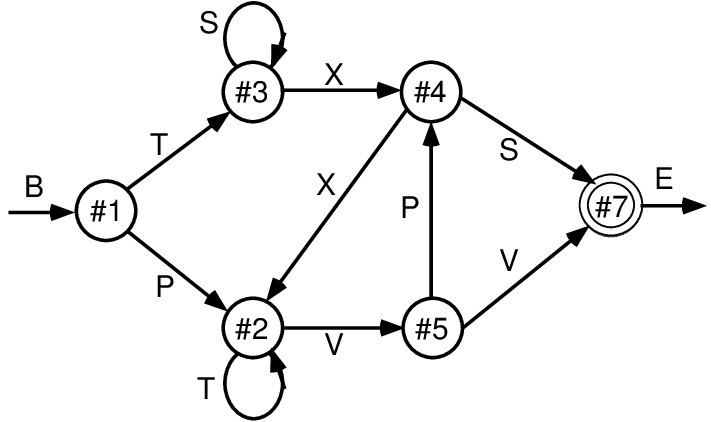}
  \caption{\footnotesize Finite state automaton (FSA) grammar used in implicit sequential learning experiments (Reber, 1967) and in early simple recurrent networks (SRNs) (Cleeremans \& McClelland, 1991).  It generates a sequence of letters according to the link transitioned between state nodes, where each outgoing link to another node has a 50\% probability of being selected.  Each letter (except for the B=begin and E=end) appears at 2 different points in the grammar, making them locally ambiguous.  This combination of randomness and ambiguity makes it challenging for a learning system to infer the true underlying structure of the grammar.}
  \label{fig.fsa_grammar}
\end{figure}

\begin{figure}
  \centering\includegraphics[width=3.5in]{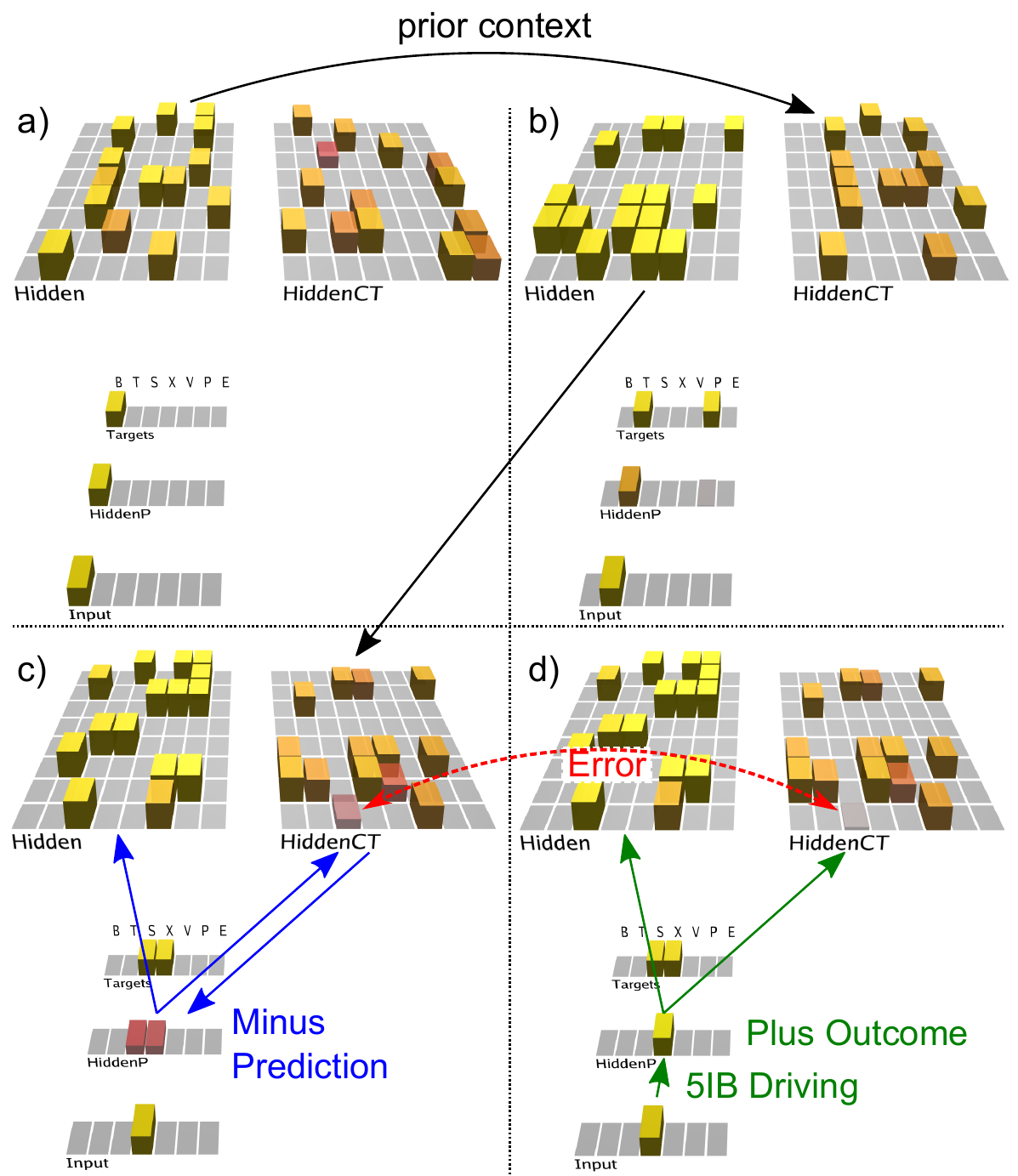}
  \caption{\footnotesize Predictive learning model applied to the FSA grammar shown in previous figure. The first three panels (a, b, c) show the prediction state (end of the \emph{minus} phase, e.g., the first 75 ms of an alpha cycle) of the trained model on the first three steps of the sequence `BTX' (plus phases also occurred, but are not shown).  The last panel (d) shows the plus phase after the third step.  The \emph{Input} layer provides the 5IB drivers for the corresponding \emph{HiddenP} pulvinar layer, so the plus phase is always based on the specific randomly-selected path taken.  The \emph{Targets} layer is purely for display, showing the two valid possible labels that could have been predicted.  To track learning, the model's prediction is scored as accurate if either or both targets are activated.  Computationally, the model is similar to an SRN, where the CT layer that drives the prediction over the pulvinar encodes the activation state from the previous time step (alpha cycle), due to the phasic bursting of the 5IB neurons that drive CT updating.  Note how the CT layer in b) reflects the Hidden activation state in a), and likewise for c) reflecting b).  This is evident because we're using one-to-one connectivity between Hidden and HiddenCT layers (which works well in general, along with full lateral connectivity within the CT layer).  Thus, even though the correct answer is always present on the Input layer for each step, the CT layer is nevertheless attempting to predict this Input based on the information from the prior time step.  {\bf a)} In the first step, the B label is unambiguous and easily predicted (based on prior E context). {\bf b)} In the 2nd step, the network correctly guesses that the T label will come next, but there is a faint activation of the other P alternative, which is also activated sometimes based on prior learning history and associated minor weight tweaks.  {\bf c)} In the 3rd step, both S and X are equally predicted.  {\bf d)} In the \emph{plus} phase, only the Input pattern (`X' on this trial) drives HiddenP activations, and the projections from pulvinar back to the cortex convey both the minus-phase prediction and plus-phase actual input.  You can see one HiddenCT neuron, just above the arrow, visibly changes its activation as a result (and all neurons experience smaller changes), and learning in all these cortical (Hidden) layer neurons is a function of their local temporal difference between minus and plus phases.}
  \label{fig.fsa_net}
\end{figure}

To illustrate and test the predictive learning abilities of this biologically based model, we first ran a classical test of sequence learning \citep{Reber67,CleeremansMcClelland91} that has been explored using simple recurrent networks (SRNs) \citep{Elman90,Jordan89}.  The biologically based model was implemented using the \emph{Leabra} algorithm, which is a comprehensive framework that uses conductance-based point neuron equations, inhibitory competition, bidirectional connectivity, and the biologically plausible temporal difference learning mechanism described above \citep{OReillyHazyHerd16,OReillyMunakataFrankEtAl12,OReillyMunakata00,OReilly98,OReilly96}.  Leabra serves as a model of the bidirectionally connected processing in the cortical superficial layers, and has been used to simulate a large number of different cognitive neuroscience phenomena.  It is described in the Appendix, which also provides a detailed mapping between the SRN and our biological model.

As shown in Figure~\ref{fig.fsa_grammar}, sequences were generated according to a finite state automaton (FSA) grammar, as used in implicit sequence learning experiments by \citet{Reber67}.  Each node has a 50\% random branching to two different other nodes, and the labels generated by node transitions are locally ambiguous (except for the B=begin and E=end states).  Thus, integration over time and across many iterations are required to infer the systematic underlying grammar.  It is a reasonably challenging task for SRNs and people to learn and provides an important validation of the power of these predictive learning mechanisms.  Given the random branching, accurately predicting the specific path taken is impossible, but we can score the model's output as correct if it activates either or both of the possible branches for each state.

The model (Figure~\ref{fig.fsa_net}) required around 20 epochs of 25 sequences through the grammar to learn it to the point of making no prediction errors for 5 epochs in a row (which guarantees that it had completely learned the task).  This model is available in the standard \emph{emergent} distribution, at \url{https://github.com/emer/leabra/tree/master/examples/deep_fsa}.  A few steps through a sequence are shown in the figure, illustrating how the CT context layer, which drives the P pulvinar layer prediction, represents the information present on the \emph{previous} alpha cycle time step.  Thus, the network is attempting to predict the current Input state, which then drives the pulvinar plus phase at the end of each alpha cycle, as shown in the last panel.  On each trial, the difference between plus and minus phases locally over each cortical neuron drives its synaptic weight changes, which accumulate over trials to allow accurate prediction of the sequences, to the extent possible given their probabilistic nature.

\section{Predictive Learning of Object Categories in IT Cortex}

\begin{figure}
  \centering\includegraphics[width=4in]{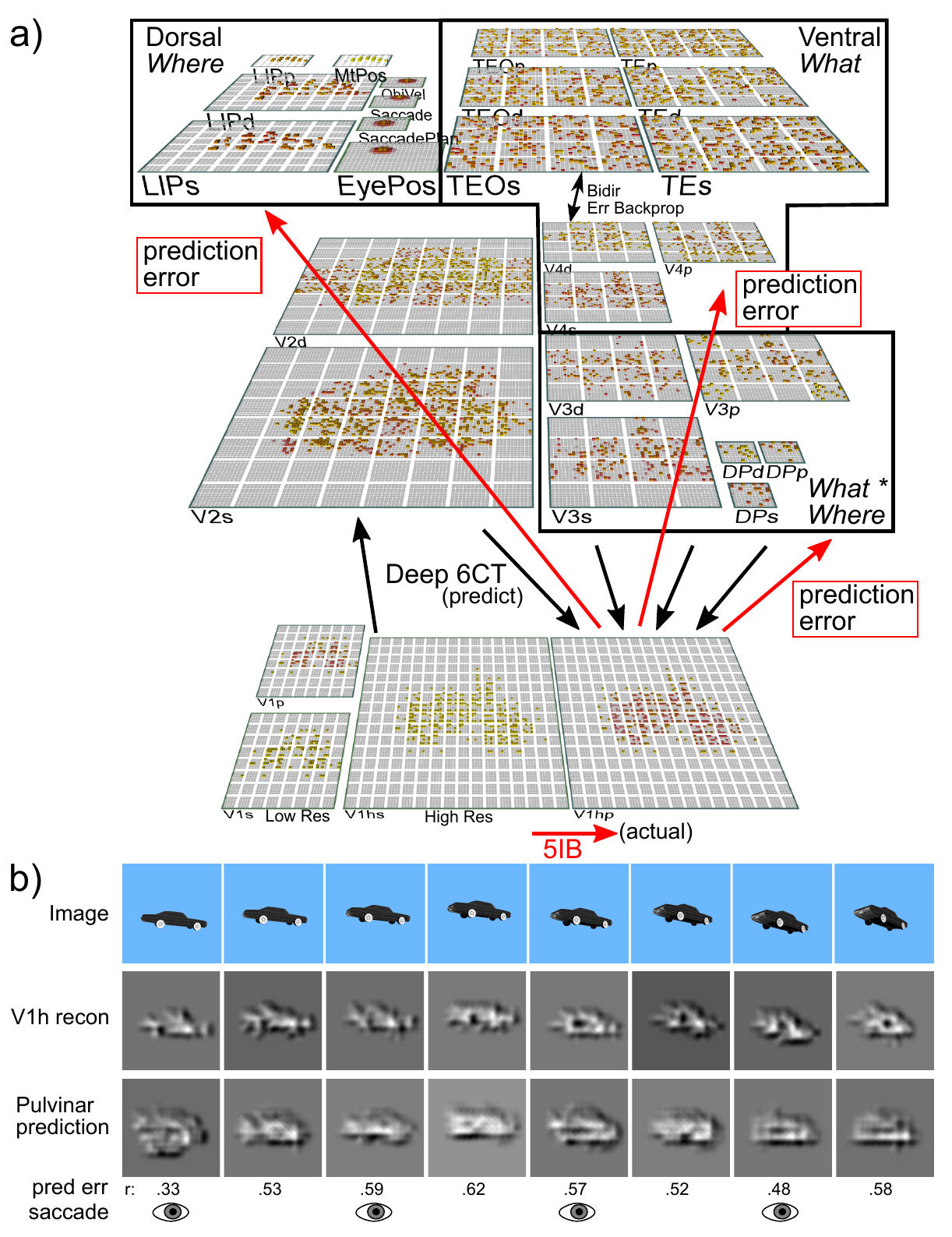}
  \caption{\footnotesize {\bf a)} The \emph{What-Where-Integration, WWI} deep predictive learning model. The dorsal \emph{Where} pathway learns first, using easily-abstracted \emph{spatial blobs}, to predict object location based on prior motion, visual motion, and saccade efferent copy signals.  This drives strong top-down inputs to lower areas with accurate spatial predictions, leaving the \emph{residual} error concentrated on \emph{What} and \emph{What * Where} integration.  The V3 and DP (dorsal prelunate) constitute the \emph{What * Where} integration pathway, binding features and locations.  V4, TEO, and TE are the \emph{What} pathway, learning abstracted object category representations, which also drive strong top-down inputs to lower areas.  Suffixes: \emph{s} = superficial, \emph{d} = deep, \emph{p} = pulvinar. {\bf c)} Example sequence of 8 alpha cycles that the model learned to predict, with the reconstruction of each image based on the V1 gabor filters (\emph{V1h recon}), and model-generated prediction (correlation $r$ prediction error shown).  The low resolution and reconstruction distortion impair visual assessment, but $r$ values are well above the $r$'s for each V1 state compared to the previous time step (mean = .38, min of .16 on frame 4 --- see Appendix for more analysis).  Eye icons indicate when a saccade occurred.}
  \label{fig.model}
\end{figure}

Now we describe a large-scale, systems-neuroscience implementation of the proposed thalamocortical predictive error-driven learning framework, in a model of visual predictive learning (Figure~\ref{fig.model}).  Our second major objective, and a critical question for predictive learning, is determining whether the model can develop high-level, abstract ways of representing the raw sensory inputs, while learning from nothing but predicting these low-level visual inputs.  We showed the model brief movies of 156 3D object exemplars drawn from 20 different basic-level categories (e.g., car, stapler, table lamp, traffic cone, etc.) selected for their overall shape diversity from the CU3D-100 dataset \citep{OReillyWyatteHerdEtAl13}.  The objects moved and rotated in 3D space over 8 movie frames, where each frame was sampled at the alpha frequency (Figure~\ref{fig.model}b).  There were also saccadic eye movements every other frame, introducing an additional, realistic, predictive-learning challenge.  An efferent copy signal enabled full prediction of the effects of the eye movement, and allows the model to capture the signature predictive remapping phenomenon \citep{DuhamelColbyGoldberg92,CavanaghHuntAfrazEtAl10,NeupaneGuittonPack17}.  The \emph{only} learning signal available to the model was the prediction error generated by the temporal difference between what it predicted to see in the V1 input in the next frame and what was actually seen.

As described in detail in the Appendix, our model was constructed to capture critical features of the visual system, including the major division between a dorsal \emph{Where} and ventral \emph{What} pathway \citep{UngerleiderMishkin82}, and the overall hierarchical organization of these pathways derived from detailed connectivity analyses \citep{RocklandPandya79,FellemanVanEssen91,MarkovVezoliChameauEtAl14,MarkovErcsey-RavaszGomesEtAl14}.  In addition to these biological constraints, we conducted extensive exploration of the connectivity and architecture space, and found a remarkable convergence between what worked functionally and the known properties of these pathways \citep{OReillyWyatteRohrlich17}.  For example, the feedforward pathway has projections from lower-level superficial layers to superficial layers of higher levels, while feedback originated in both the superficial and deep and projected back to both \citep{RocklandPandya79,FellemanVanEssen91}.  Also, consistent with the core features of the pulvinar pathways discussed above, deep layer predictive (6CT) inputs originated in higher levels, while driver (5IB) inputs originated in lower levels.  For simplicity we organized the model layers in terms of these driver inputs, whereas the topographic organization of pulvinar in the brain is organized more according to the 6CT projection loops \citep{Shipp03}.

Another important set of parameters are the strength of deep-layer recurrent projections, which influence the timescale of temporal integration, producing a simple biologically based version of \emph{slow feature analysis} \citep{WiskottSejnowski02,Foldiak91}. We followed the biological data suggesting that recurrence increases progressively up the visual hierarchy \citep{ChaudhuriKnoblauchGarielEtAl15}.  It was essential that the \emph{Where} pathway learn first, consistent with extant data \citep{BourneRosa06,KiorpesPriceHall-HaroEtAl12}, including early pathways interconnecting LIP and pulvinar \citep{BridgeLeopoldBourne16}, and a rare asymmetric pathway, from V1 to LIP \citep{MarkovErcsey-RavaszGomesEtAl14}, providing a direct short-cut for high-level spatial representations in LIP.  Results from various informative model architecture and parameter manipulations are discussed below after the primary results from the standard intact model.  Learning curves and other model details are shown in the Appendix.  We have also implemented a full \emph{de-novo} replication of the model in a new modeling framework, which also replicated the results shown here (see Appendix for more details).  Furthermore, much of the model was originally developed in the context of a set of object-like patterns generated systematically from a set of simple line features, indicating the general applicability of this architecture.

\begin{figure}
  \centering\includegraphics[width=2in]{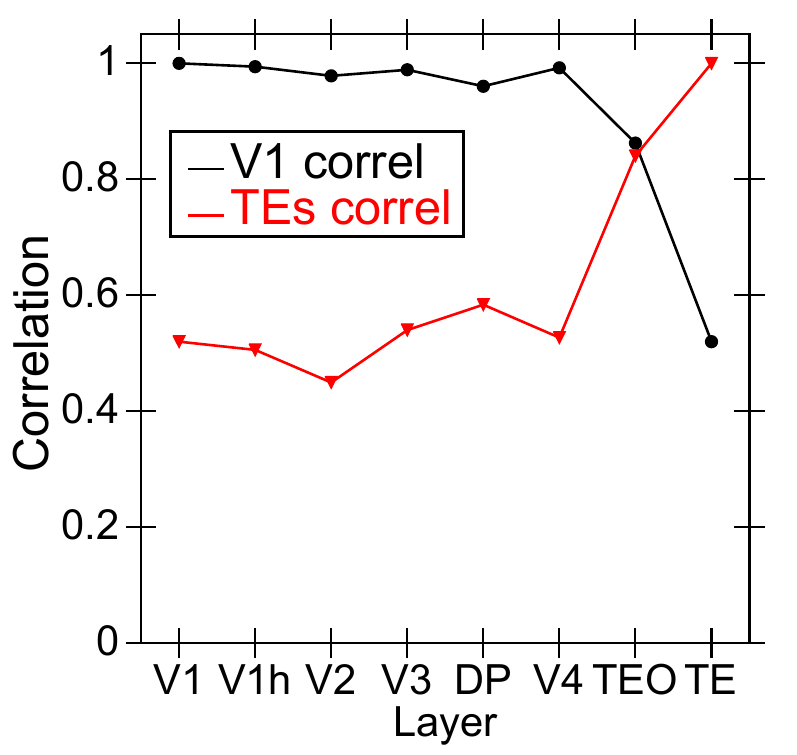}
  \caption{\footnotesize Emergence of abstract category structure over the hierarchy of layers, comparing similarity structure in each layer vs that present in V1 (black line) or in TE (red line).  Both cases, which are roughly symmetric, clearly show that IT layers (TEO, TE) progressively differentiate from raw input similarity structure present in V1, and, critically, that the model has learned structure beyond that present in the input.  This is is the simplest, most objective summary statistic showing this progressive emergence of structure, while subsequent figures provide a more concrete sense of what kinds of representations actually developed.}
  \label{fig.simat-lays}
\end{figure}

To directly address the question of whether the hierarchical structure of the network supports the development of abstract, higher-level representations that go beyond the information present in the visual inputs, we applied a second-order similarity measure across the object-level similarity matrices computed at each layer in the network (Figure~\ref{fig.simat-lays}).  This shows the extent to which the similarity matrix across objects in one layer is itself similar to the object similarity matrix in another layer, in terms of a correlation measure across these similarity matrices.  Critically, this measure does not depend on any kind of subjective interpretation of the learned representations --- it just tells us whether whatever similarity structure was learned differs across the layers.  Starting from either V1 compared to all higher layers, or the highest TE layer compared to all lower layers, we found a consistent pattern of progressive emergence of the object categorization structure in the upper IT pathway (TEO, TE).

This analysis confirms that indeed the IT category structure is significantly different from that present at the level of the V1 primary visual input.  Thus the model, despite being trained only to generate accurate visual input-level predictions, has learned to represent these objects in an abstract way that goes beyond the raw input-level information.  We further verified that at the highest IT levels in the model, a consistent, spatially-invariant representation is present across different views of the same object (e.g., the average correlation across frames within an object was .901).

\begin{figure}
  \centering\includegraphics[width=6in]{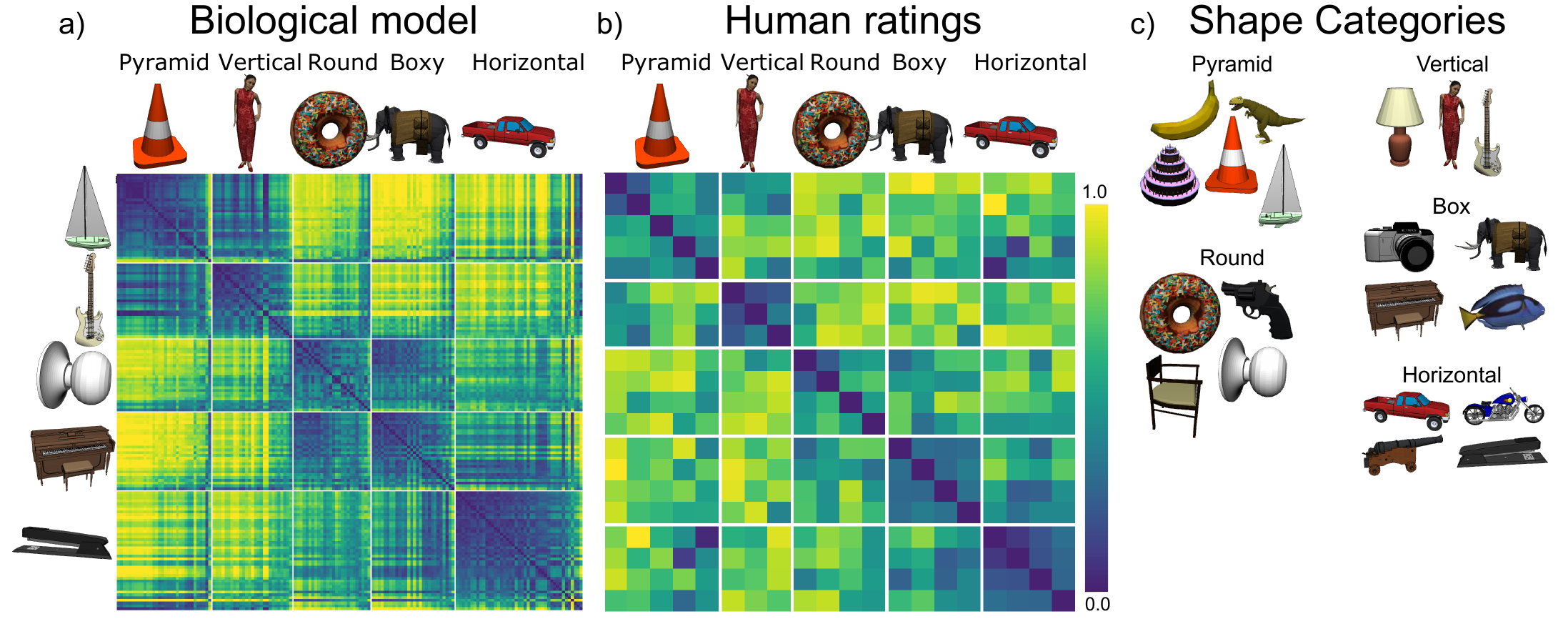}
  \caption{\footnotesize {\bf a)} Category similarity structure that emerged in the highest layer, TE, of the biologically based predictive learning model, showing \emph{dissimilarity} (1-correlation) of the TE representation for each 3D object against every other 3D object (156 total objects).  Blue cells have high similarity.  Model has learned block-diagonal clusters or categories of high-similarity groupings, contrasted against dissimilar off-diagonal other categories.  Clustering maximized average \emph{within -- between} dissimilarity (see Appendix), and clearly corresponded to the shown shape-based categories, with exemplars from each category shown.  Also, all items from the same basic-level object categories (N=20) are reliably subsumed within learned categories. {\bf b)} Human similarity ratings for the same 3D objects, presented with the V1 reconstruction (see Fig 1b) to capture coarse perception in the model, aggregated by 20 basic-level categories (156 x 156 matrix was too large to sample densely experimentally).  Each cell is 1 - proportion of time given object pair was rated more similar than another pair (see Appendix).  The human matrix shares the same centroid categorical structure as the model (confirmed by permutation testing and agglomorative cluster analysis, see Appendix), indicating that human raters used the same shape-based category structure. {\bf c)} One object from each of the 20 basic level categories, organized into the shape-based categories.  The Vertical, Box and Horizontal categories are fairly self-evident and the model was most consistent in distinguishing those, along with subsets of the Pyramid (layer-cake, traffic-cone, sailbot) and Round (donut, doorknob) categories, while banana, trex, chair, and handgun were more variable.}
  \label{fig.rsa}
\end{figure}

To better understand the nature of these learned representations, Figure~\ref{fig.rsa} shows a representational similarity analysis (RSA) on the activity patterns at each layer in the model, which reveals the explicit categorical structure of the learned representations \citep{KriegeskorteMurBandettini08,CadieuHongYaminsEtAl14}.  As shown in Figure~\ref{fig.rsa}a, we found that the highest IT layer (TE) produced a systematic organization of the 156 3D objects into 5 categories.  In our admittedly subjective judgment, these categories seemed to correspond to the overall shape of the objects, as shown by the object exemplars in the figure (pyramid-shaped, vertically-elongated, round, boxy / square, and horizontally-elongated).  Furthermore, the basic-level categories were subsumed within these broader shape-level categories, so the model appears to be sensitive to the coherence of these basic-level categories as well, but apparently their shapes were not sufficiently distinct between categories to drive differentiated TE-level representations for each such basic-level category.

Given that the model only learns from a passive visual experience of the objects, it has no access to any of the richer interactive multi-modal information that people and animals would have.  Furthermore, as evident in Figure~\ref{fig.model}b, the relatively low resolution of the V1 layers (required to make the model tractable computationally) means that complex visual details are not reliably encoded (and even so, are not generally reliable across object exemplars), such that the overall object shape is the most salient and sensible basis for categorization for this model.

Although these object shape categories appeared sensible to us, we ran a simple experiment to test whether a sample of 30 human participants would use the same category structure in evaluating the pairwise similarity of these objects.  Figure~\ref{fig.rsa}b shows the results, confirming that indeed this same organization of the objects emerged in their similarity judgments.  These judgments were based on the V1 reconstruction as shown in Figure~\ref{fig.model}b to capture the model's coarse-grained perception; see Appendix for methods and further analysis.

\begin{figure}
  \centering\includegraphics[width=4in]{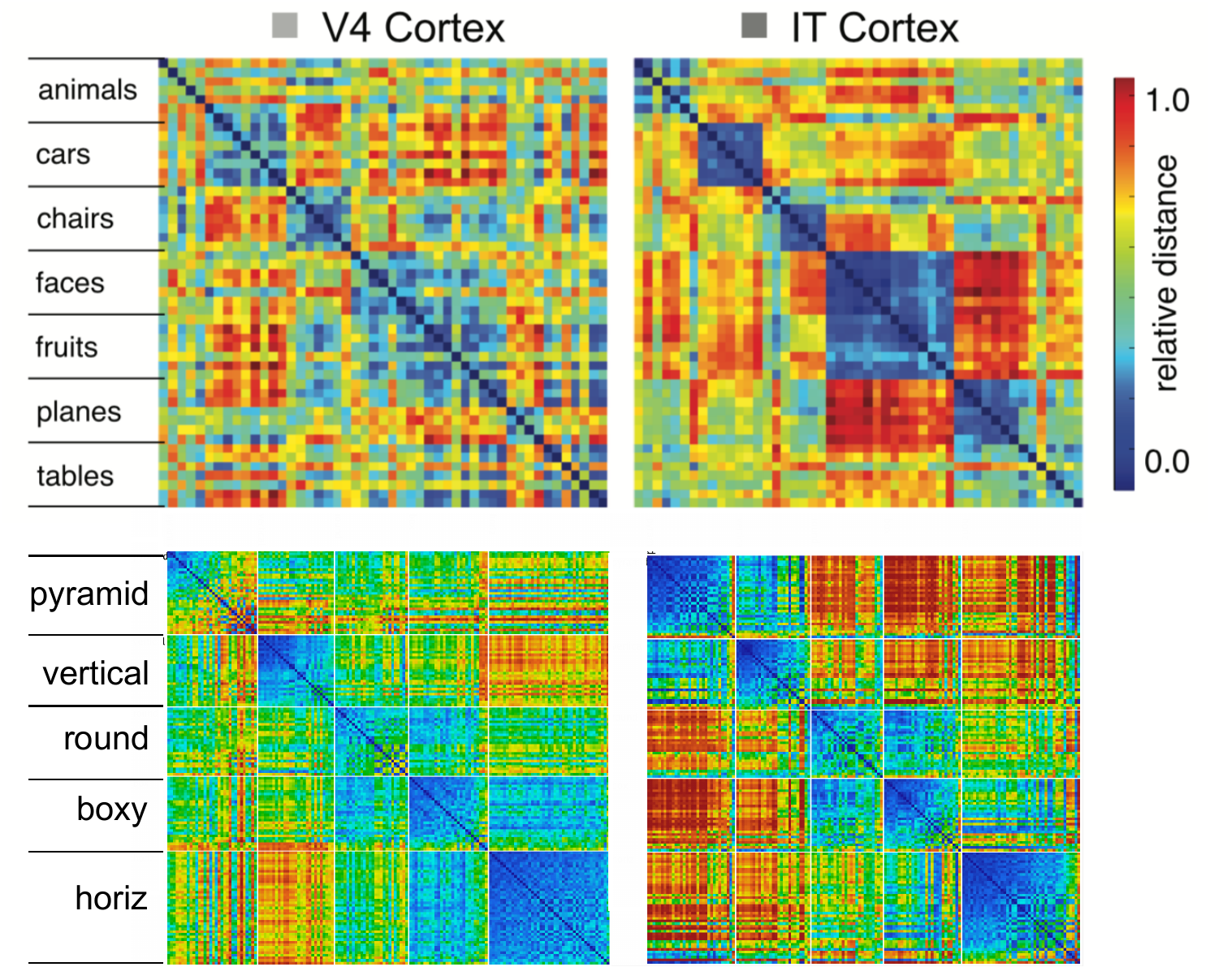}
  \caption{\footnotesize Comparison of progression from V4 to IT in macaque monkey visual cortex (top row, from Cadieu et al., 2014) versus same progression in model (replotted using comparable color scale).  Although the underlying categories are different, and the monkeys have a much richer multi-modal experience of the world to reinforce categories such as foods and faces, the model nevertheless shows a similar qualitative progression of stronger categorical structure in IT, where the block-diagonal highly similar representations are more consistent across categories, and the off-diagonal differences are stronger and more consistent as well (i.e., categories are also more clearly differentiated).  Note that the critical difference in our model versus those compared in Cadieu et al. 2014 and related papers is that they explicitly trained their models on category labels, whereas our model is \emph{entirely self-organizing} and has no external categorical training signal.}
  \label{fig.macaque}
\end{figure}

The progressive emergence of increasingly abstract category structure across visual areas, evident in Figure~\ref{fig.simat-lays}, has been investigated in recent comparisons between monkey electrophysiological recordings and deep convolutional neural networks (DCNNs), which provide a reasonably good fit the the overall progressive pattern of increasingly categorical organization \citep{CadieuHongYaminsEtAl14}.  However, these DCNNs were trained on large datasets of human-labeled object categories, and it is perhaps not too surprising that the higher layers closer to these category output labels exhibited a greater degree of categorical organization.  In contrast, because the only source of learning in our model comes from prediction errors over the V1 input layers, the graded emergence of an object hierarchy here reflects a truly self-organizing learning process.

Figure~\ref{fig.macaque} compares the similarity structures in layers V4 and IT in macaque monkeys \citep{CadieuHongYaminsEtAl14} with those in corresponding layers in our model.  In both the monkeys and our model, the higher IT layer builds upon and clarifies the noisier structure that is emerging in the earlier V4 layer, showing that our model replicates the essential qualitative hierarchical progression in the brain.  As noted, we would not expect our model to exactly replicate the detailed object-specific similarity structure found in macaques, due to the impoverished nature of our model's experience, so this comparison remains qualitative in terms of the respective differences between V4 and IT in each model, rather than a direct comparison of the similarity structure between corresponding layers in the model and the macaque.  In the future, when we can scale up our model and tune the attentional processing dynamics necessary to deal with cluttered visual scenes, we will be able to train our model on the same images presented to the macaques, and can provide this more direct comparison.

Finally, we did not use analyses based on decoding techniques, because with high-dimensional distributed neural representations, it is generally possible to decode many different features that are not otherwise compactly and directly represented \citep{FusiMillerRigotti16}.  In preliminary work using decoding in the context of the simpler feature-based input patterns, we indeed found that decoding was not a very sensitive measure of the differentiation of representations across layers, which is so clearly evident in Figure~\ref{fig.simat-lays}.  Thus, as advocates of the RSA approach have argued, measuring similarity structure evident in the activity patterns over a given layer generally provides a clearer picture of what that layer is explicitly encoding \citep{KriegeskorteMurBandettini08}.

In summary, the model learned an abstract category organization that reflects the overall visual shapes of the objects as judged by human participants, in a way that is invariant to the differences in motion, rotation, and scaling that are present in the V1 visual inputs.  We are not aware of any other model that has accomplished this signature computation of the ventral \emph{What} pathway in a purely self-organizing manner operating on realistic 3D visual objects, without any explicit supervised category labels.  Furthermore, our model does this using a learning algorithm directly based on detailed properties of the underlying biological circuits in this pathway, providing a coherent overall account.

\subsection{Backpropagation Comparison Models}

\begin{figure}
  \centering\includegraphics[width=3.5in]{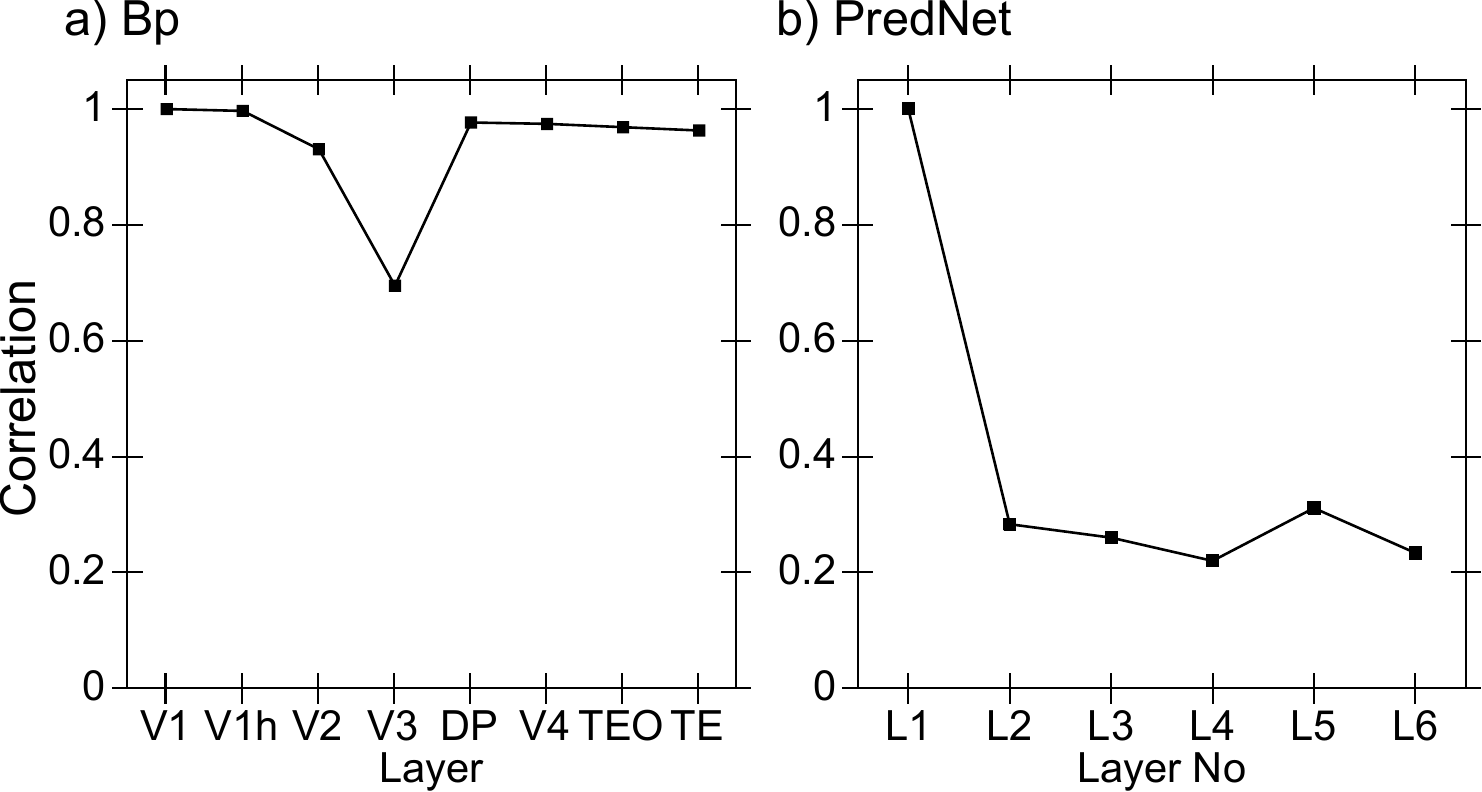}
  \caption{\footnotesize Similarity of similarity structure across layers for the comparison backprop models, comparing each layer to the first layer.  {\bf a)} Backpropagation (Bp) model with the same What / Where structure as the biological model.  Unlike the biologically based model (Figure~\ref{fig.simat-lays}) the higher IT layers (TE, TEO) do not diverge significantly from the similarity structure present in V1, indicating that the model has not developed abstractions beyond the structure present in the visual input.  Layer V3 is most directly influenced by spatial prediction errors, so it differs from both in strongly encoding position information.  {\bf b)} PredNet model, which has 6 layers.  Layers 2-6 diverge from layer 1, but there is no progressive change in the higher layers as we see in our model moving from V4 to TEO. The divergence in correlation starting at layer 2 is likely due to the fact that higher layers only encode errors, not stimulus-driven positive representations of the input.  Aside from this large distinction (which is inconsistent with the similarity in neural coding seen in actual V1 and V2 recordings), there is no evidence of a cumulative development of abstraction in higher layers.}
  \label{fig.bpred-v1sim}
\end{figure}

\begin{figure}
  \centering\includegraphics[width=3.5in]{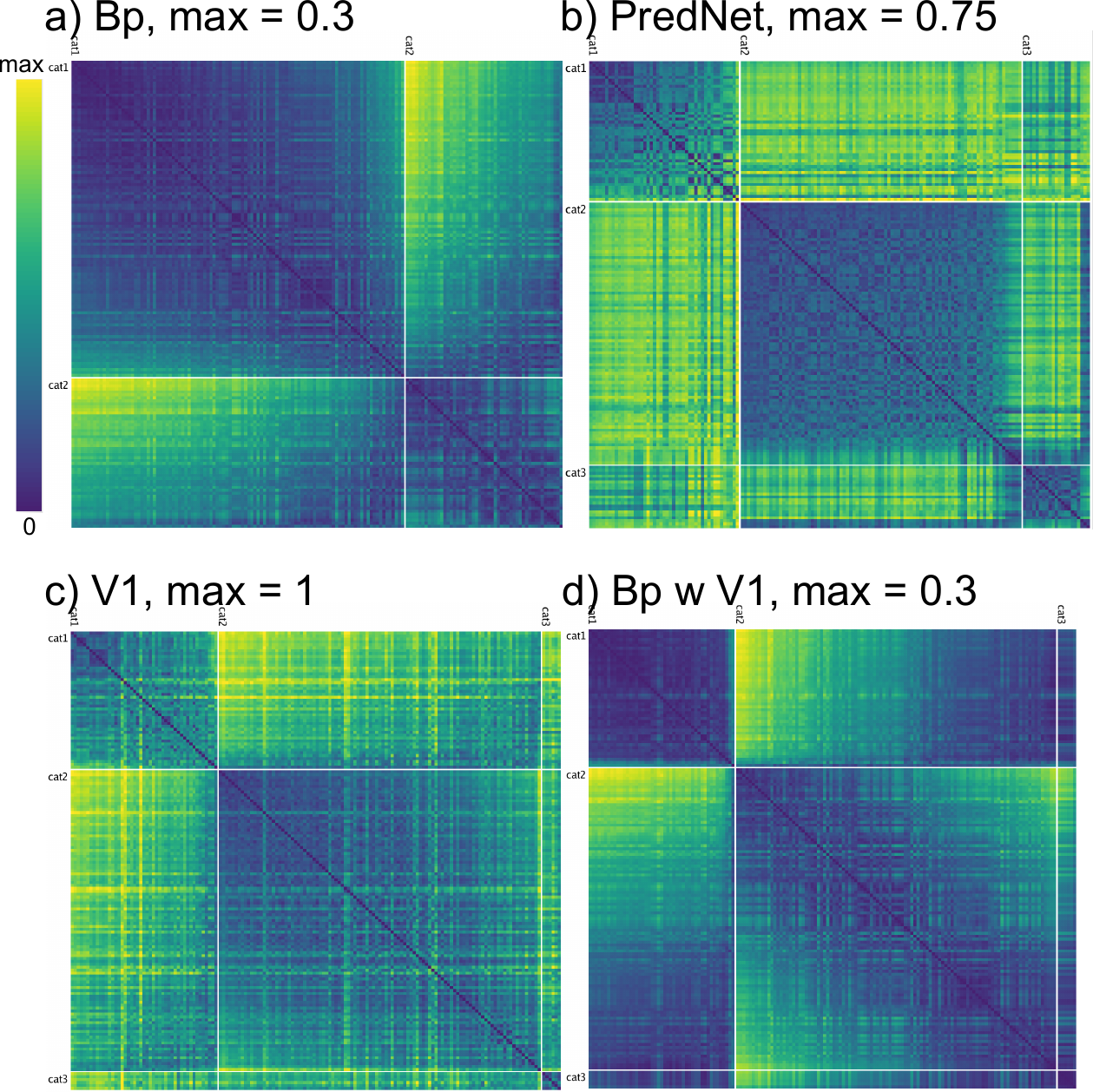}
  \caption{\footnotesize {\bf a)} Best-fitting category similarity for TE layer of the backpropagation (Bp) model with the same What / Where structure as the biological model.  Only two broad categories are evident, and the lower \emph{max} distance (0.3 vs. 1.5 in biological model) means that the patterns are much less differentiated overall.  {\bf b)} Best-fitting similarity structure for the PredNet model, in the highest of its layers (layer 6), which is more differentiated than Bp (max = 0.75) but also less cleanly similar within categories (i.e., less solidly blue along the block diagonal), and overall follows a broad category structure similar to V1. {\bf c)} The best fitting V1 structure, which has 2 broad categories and banana is in a third category by itself.  The lack of dark blue on the block diagonal indicates that these categories are relatively weak, and every item is fairly dissimilar from every other.  {\bf d)} The Bp TE similarity values from panel a shown in the same ordering as V1 from panel c, demonstrating how the similarity structure has not diverged very much, consistent with the results shown in Figure~\ref{fig.bpred-v1sim} --- the within -- between contrast differences are 0.0838 for panel a and 0.0513 for d --- see Appendix for details.}
  \label{fig.bpred}
\end{figure}

To help discern some of the factors that contribute to the categorical learning in our model, and provide a comparison with more widely-used error backpropagation models, we tested a backpropagation-based (Bp) version of the same \emph{What} vs. \emph{Where} architecture as our biologically based predictive error model, and we also tested a standard \emph{PredNet} model \citep{LotterKreimanCox16} with extensive hyperparameter optimization (see Appendix).  Due to the constraints of backpropagation, we had to eliminate any bidirectional connectivity loops in the Bp version, but we were able to retain a form of predictive learning by configuring the V1p pulvinar layer as the final target output layer, with the target being the next visual input relative to the current V1 inputs.

Figure~\ref{fig.bpred-v1sim} shows the same second-order similarity analysis as Figure~\ref{fig.simat-lays}, to determine the extent to which these comparison networks also developed more abstract representations in the higher layers that diverge from the similarity structure present in the lowest layers.  According to this simple objective analysis, they did not --- the higher layers showed no significant, progressive divergence in their similarity structure.  The PredNet model did show a larger difference between the first layer and the rest of the layers, due to the subsequent layers encoding errors while the first layer has a positive representation of the image, but there was no progressive difference beyond that up into the higher layers.

Next, we examined the RSA matrices for the highest (TE) layer in the comparison models, also in comparison with the same for the V1 layer (Figure~\ref{fig.bpred}).  This shows that the TE layer in the Bp model formed a simple binary category structure overall, which is similar to the RSA for the V1 input layer.  It is also important to emphasize that the scales on these figures are different (as shown in their headers), such that these comparison models had much less differentiated representations overall.  Similar results were found in the PredNet model.  Because existing work with these models has typically relied on additional supervised learning and decoder-based analyses (which are essentially equivalent to an additional layer of supervised learning), these RSA-based analyses provide an important, more sensitive way of determining what they learn purely through predictive learning.

These results show that the additional biologically derived properties in our model are playing a critical role in the development of abstract categorical representations that go beyond the raw visual inputs. These properties include: excitatory bidirectional connections, inhibitory competition, and an additional Hebbian form of learning that serves as a regularizer (similar to weight decay) on top of predictive error-driven learning \citep{OReilly98,OReillyMunakata00}.  Each of these properties could promote the formation of categorical representations. Bidirectional connections enable top-down signals to consistently shape lower-level representations, creating significant attractor dynamics that cause the entire network to settle into discrete categorical attractor states.  Another indication of the importance of bidirectional connections is that a greedy layer-wise pretraining scheme, consistent with a putative developmental cascade of learning from the sensory periphery on up \citep{ShragerJohnson96,BengioYaoAlainEtAl13,Valpola14,HintonSalakhutdinov06}, did not work in our model. Instead, we found it essential that higher layers, with their ability to form more abstract, invariant representations, interact and shape learning in lower layers right from the beginning.

Furthermore, the recurrent connections within the TEO and TE layers likely play an important role by biasing the temporal dynamics toward longer persistence \citep{ChaudhuriKnoblauchGarielEtAl15}.  By contrast, backpropagation networks typically lack these kinds of attractor dynamics, and this could contribute significantly to their relative lack of categorical learning.  Hebbian learning drives the formation of representations that encode the principal components of activity correlations over time, which can help more categorical representations coalesce (and results below already indicate its importance).  Inhibition, especially in combination with Hebbian learning, drives representations to specialize on more specific subsets of the space.

Ongoing work is attempting to determine which of these is essential in this case (perhaps all of them) by systematically introducing some of these properties into the backpropagation model, though this is difficult because full bidirectional recurrent activity propagation, which is essential for conveying error signals top-down in the biological network, is incompatible with the standard efficient form of error backpropagation, and requires significantly more computationally intensive and unstable forms of fully recurrent backpropagation \citep{WilliamsZipser92,Pineda87}.  Furthermore, Hebbian learning requires dynamic inhibitory competition which is difficult to incorporate within the backpropagation framework.

\subsection{Architecture and Parameter Manipulations}

\begin{figure}
  \centering\includegraphics[width=5in]{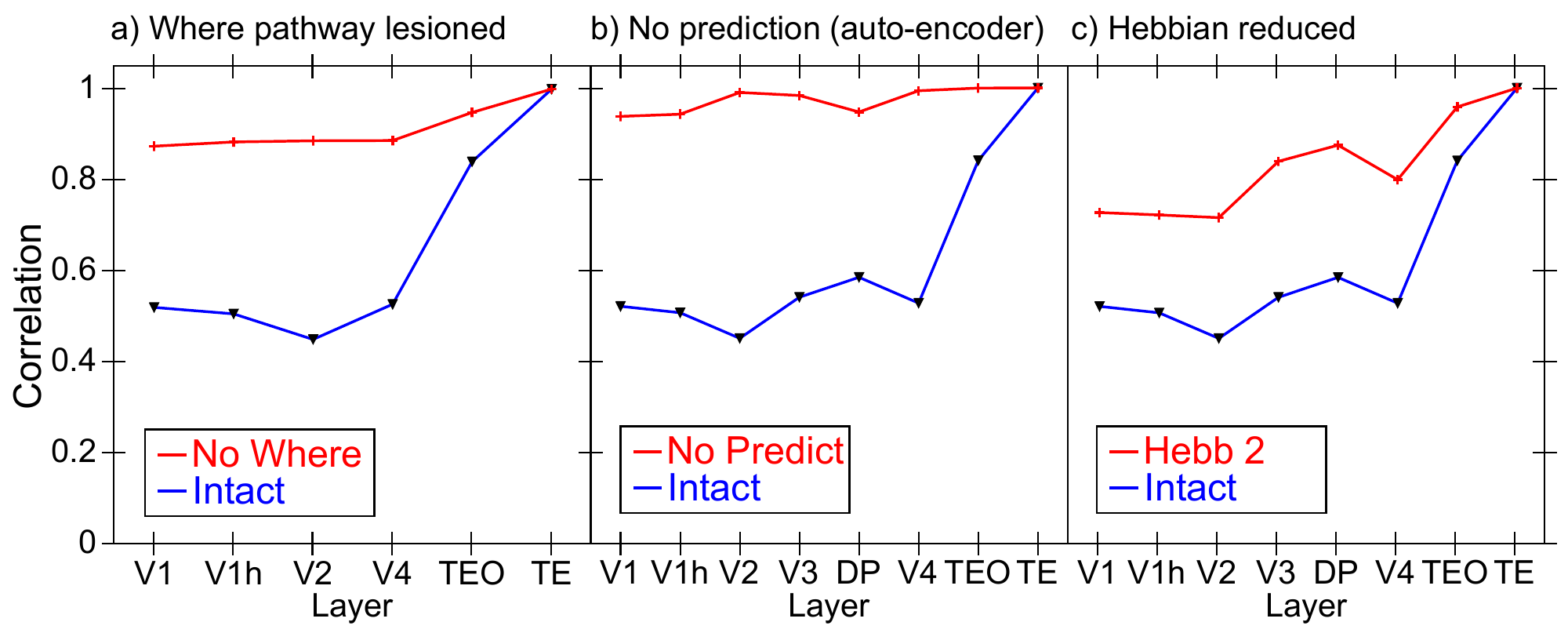}
  \caption{\footnotesize Effects of various manipulations on the extent to which TE representations differentiate from V1.  For all plots, \emph{Intact} is the same result shown in Figure~\ref{fig.simat-lays} from the intact model for ease of comparison (panel a is missing V3 and DP dorsal pathway layers).  All of the following  manipulations significantly impair the development of abstract TE categorical representations (i.e., TE is more similar to V1 and the other layers).  {\bf a)} Dorsal \emph{Where} pathway lesions, including lateral inferior parietal sulcus (LIP), V3, and dorsal prelunate (DP).  This pathway is essential for regressing out location-based prediction errors, so that the residual errors concentrate feature-encoding errors that train the \emph{What} pathway.  {\bf b)} Allowing the deep layers full access to current-time information, thus effectively eliminating the prediction demand and turning the network into an auto-encoder, which significantly impairs representation development, and supports the importance of the challenge of predictive learning for developing deeper, more abstract representations.  {\bf c)} Reducing the strength of Hebbian learning by 20\% (from 2.5 to 2), demonstrating the essential role played by this form of learning on shaping categorical representations.  Eliminating Hebbian learning entirely (not shown) prevented the model from learning anything at all, as it also plays a critical regularization and shaping role on learning.}
  \label{fig.manips}
\end{figure}

Figure~\ref{fig.manips} shows just a few of the large number of parameter manipulations that have been conducted to develop and test the final architecture.  For example, we hypothesized that separating the overall prediction problem between a spatial \emph{Where} vs. non-spatial \emph{What} pathway \citep{UngerleiderMishkin82,GoodaleMilner92}, would strongly benefit the formation of more abstract, categorical object representations in the \emph{What} pathway.  Specifically, the \emph{Where} pathway can learn relatively quickly to predict the overall spatial trajectory of the object (and anticipate the effects of saccades), and thus effectively regress out that component of the overall prediction error, leaving the residual error concentrated in object feature information, which can train the ventral \emph{What} pathway to develop abstract visual categories.  Figure~\ref{fig.manips}a shows that, indeed, when the \emph{Where} pathway is lesioned, the formation of abstract categorical representations in the intact \emph{What} pathway is significantly impaired.  We also hypothesized that full predictive learning (about the future), as compared to just encoding and decoding the current state (i.e., an auto-encoder, which is much easier computationally), is also critical for the formation of abstract categorical representations --- prediction is a ``desirable difficulty'' \citep{Bjork94}.  Figure~\ref{fig.manips}b shows that this was the case.  Finally, consistent with our hypothesis that Hebbian learning provides an important bias on learning, Figure~\ref{fig.manips}c shows the impairment associated with reducing this learning bias.

\subsection{Predictive Behavior}

\begin{figure}
  \centering\includegraphics[width=4in]{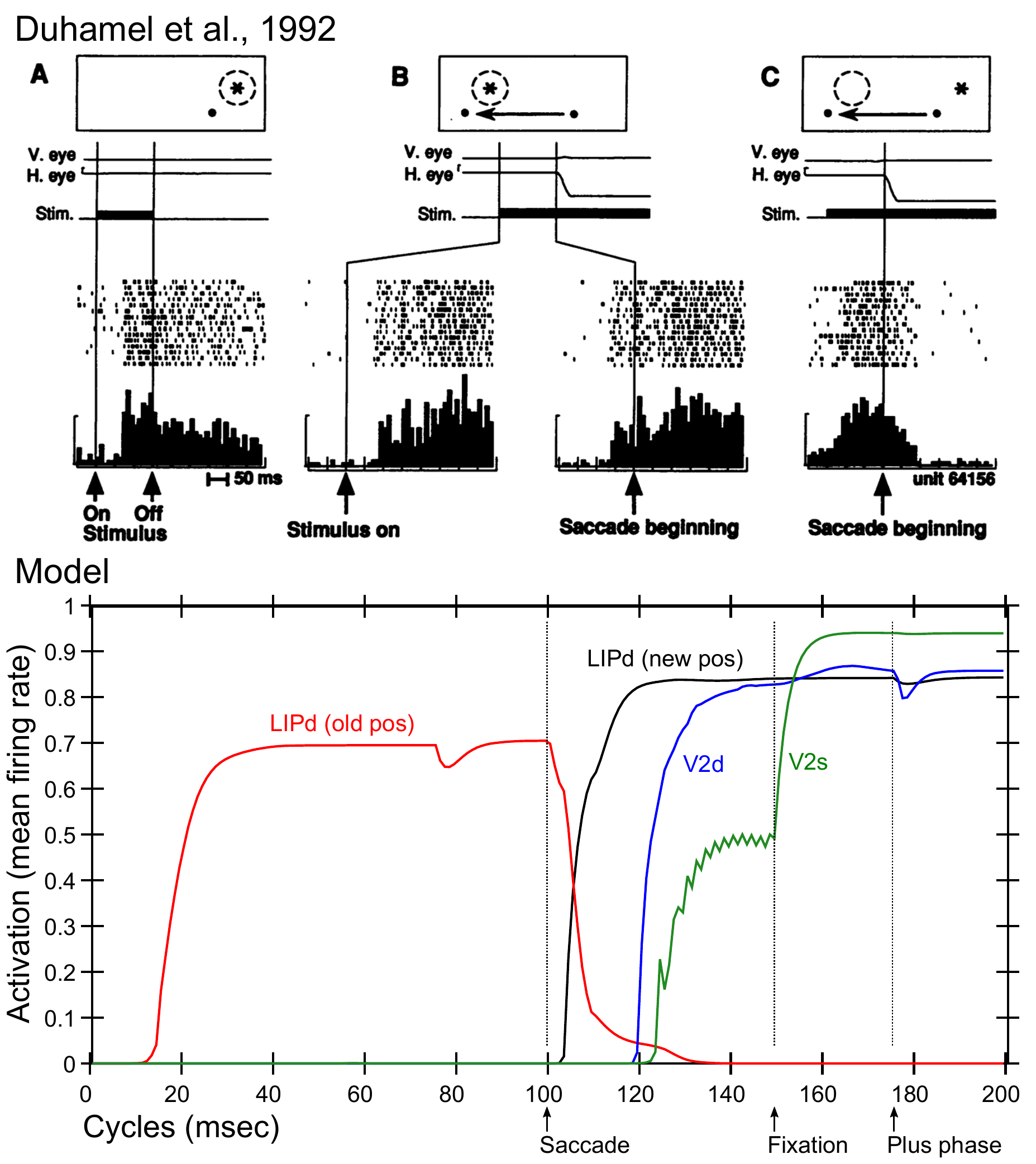}
  \caption{\footnotesize Predictive Remapping.  {\bf top:} Original remapping data in LIP from Duhamel et al (1992).  A) shows stimulus (star) response within receptive field (dashed circle) relative to fixation dot (upper right of fixation).  B) Just prior to monkey making a saccade to new fixation (moving left), stimulus is turned on in receptive field location that \emph{will be} upper right of the new fixation point, and the LIP neuron responds to that stimulus in advance of the saccade completing.  The neuron does not respond to the stimulus in that location if it is not about to make a saccade that puts it within its receptive field (not shown).  This is predictive remapping.  C) response to the old stimulus location goes away as saccade is initiated.  {\bf bottom:} Data from our model, from individual units in LIPd, V2d, and V2s, showing that the LIP deep neurons respond to the saccade first, activating in the new location and deactivating in the old, and this LIP activation goes top-down to V3 and V2 to drive updating there, generally at a longer latency and with less activation especially in the superficial layers.  When the new stimulus appears at the point of fixation (after a 50 ms saccade here), the \emph{primed} V2s units get fully activated by the incoming stimulus.  But the deep neurons are insulated from this superficial input until the plus phase, when the cascade of 5IB firing drives activation of the actual stimulus location into the pulvinar, which then reflects up into all the other layers.}
  \label{fig.remap_units}
\end{figure}

A signature example of predictive behavior at the neural level in the brain is the \emph{predictive remapping} of visual space in anticipation of a saccadic eye movements \citep{DuhamelColbyGoldberg92,ColbyDuhamelGoldberg97,GottliebKusunokiGoldberg98,NakamuraColby02,MarinoMazer16} (Figure~\ref{fig.remap_units}a).  Here, parietal neurons start to fire at the \emph{future} receptive field location where a currently-visible stimulus will appear after a planned saccade is actually executed. Remapping has also been shown for border ownership neurons in V2 \citep{OHerronHeydt13} and in area V4 \citep{NeupaneGuittonPack16,NeupaneGuittonPack20}. These are examples, we believe, of a predictive process operating throughout the neocortex to predict what will be experienced next.  A major consequence of this predictive process is the perception of a stable, coherent visual world despite constant saccades and other sources of visual change.

Figure~\ref{fig.remap_units}b shows that our model exhibits this predictive remapping phenomenon.  Specifically, LIP, which is most directly interconnected with the saccade efferent copy signals, is the first to predict the new location, and it then drives top-down activation of lower layers.  This top-down dynamic is consistent with the account of predictive remapping given by \citet{Wurtz08} and \citet{CavanaghHuntAfrazEtAl10}, who argue that the key remapping takes place at the high levels of the dorsal stream, which then drive top-down activation of the predicted location in lower areas, instead of the alternative where lower-levels remap themselves based on saccade-related signals.  The lower-level visual layers are simply too large and distributed to be able to remap across the relevant degrees of visual angle --- the extensive lateral connectivity needed to communicate across these areas would be prohibitive.

\section{Neural Data and Predictions}

Having tested the computational and functional learning properties of this biologically based predictive learning mechanism, we now return to consider some of the most important neural data of relevance to our hypotheses, beyond that summarized in the introduction, including contrasts with a widely-discussed alternative framework for predictive coding, and some of the extensive data on alpha frequency effects, followed by a discussion of predictions that would clearly test the validity of this framework.

\subsection{Additional Neuroscience Data}

We begin with data relevant to the basic neural-level properties of the framework.  First, a central element of the proposed model is the alpha cycle bursting, and subsequent inter-burst pauses, in the 5IB neurons.  Direct electrophysiological recording of deep layer neurons shows periodic alpha-scale bursting for continuous tones \emph{in awake animals} \citep{LuczakBarthoHarris09,LuczakBarthoHarris13,SakataHarris09,SakataHarris12}.  \emph{In vitro}, a variety of potential mechanisms behind the generation and synchronization of the 5IB bursts driving this alpha cycle have been identified \citep{ConnorsGutnickPrince82,SilvaAmitaiConnors91,FranceschettiGuatteoPanzicaEtAl95}.  Furthermore, the pulvinar has been shown to drive alpha-frequency synchronization of cortical activity across areas in the alpha band in awake behaving animals \citep{SaalmannPinskWangEtAl12}.  We review the larger alpha frequency literature in more detail below, but it is critical to emphasize that this alpha bursting dynamic is actually found in awake, behaving animals, because so many other bursting and up / down state phenomena have recently been shown to only occur in anesthetized brains, including bursting in the thalamic TRC neurons.

In contrast to the 5IB bursting, the 6CT neurons exhibit regular spiking behavior,  \citep{Thomson10,ThomsonLamy07}, providing consistent activation to the pulvinar. Also, they do not have axonal branches that project to other cortical areas --- the subpopulation that projects to the pulvinar only project there and not to other cortical areas \citep{PetrofViaeneSherman12}, whereas there are other layer 6 neurons that do project to other cortical areas.  This distinct connectivity is consistent with a specific role of this neuron type in generating predictions in the pulvinar.   The 6CT synaptic inputs on pulvinar TRCs have metabatropic glutamate receptors (mGluR) that have longer time-scale temporal dynamics consistent with the alpha period (100 ms) and even longer \citep{Sherman14}, and the 6CT neurons themselves also have temporally-delayed responding \citep{HarrisShepherd15,SakataHarris09,Thomson10}.  Furthermore, they have significantly more plasticity-inducing NMDA receptors compared to the 5IB projections \citep{UsreySherman18}.  These properties are consistent with the 6CT inputs driving a longer-integrated prediction signal that is subject to learning, whereas the 5IB are likely non-plastic and their effects are tightly localized in time.

The 5IB inputs often have distinctive \emph{glomeruli} structures at their synapses onto pulvinar neurons, which contain a complete feedforward inhibition circuit involving a local inhibitory interneuron, in addition to the direct strong excitatory driver input \citep{WilsonBoseShermanEtAl84}.  Computationally, this can provide a balanced level of excitatory and inhibitory drive so as to not overly excite the receiving neuron, while still dominating its firing behavior.

Although there are well-documented and widely-discussed burst vs. tonic firing modes in pulvinar neurons \citep{ShermanGuillery06}, there is not much evidence of these playing a clear role in the awake, behaving state, and as noted earlier the growing electrophysiological evidence shows a remarkable correspondence between cortical and pulvinar response properties across multiple different pulvinar areas in this awake state.  Nevertheless, there may be important dynamics arising from these firing modes that are more subtle or emerge in particular types of state transitions that may have yet to be identified.

\subsection{Contrast with Explict Error (EE) Frameworks}

\begin{figure}
  \centering\includegraphics[width=4in]{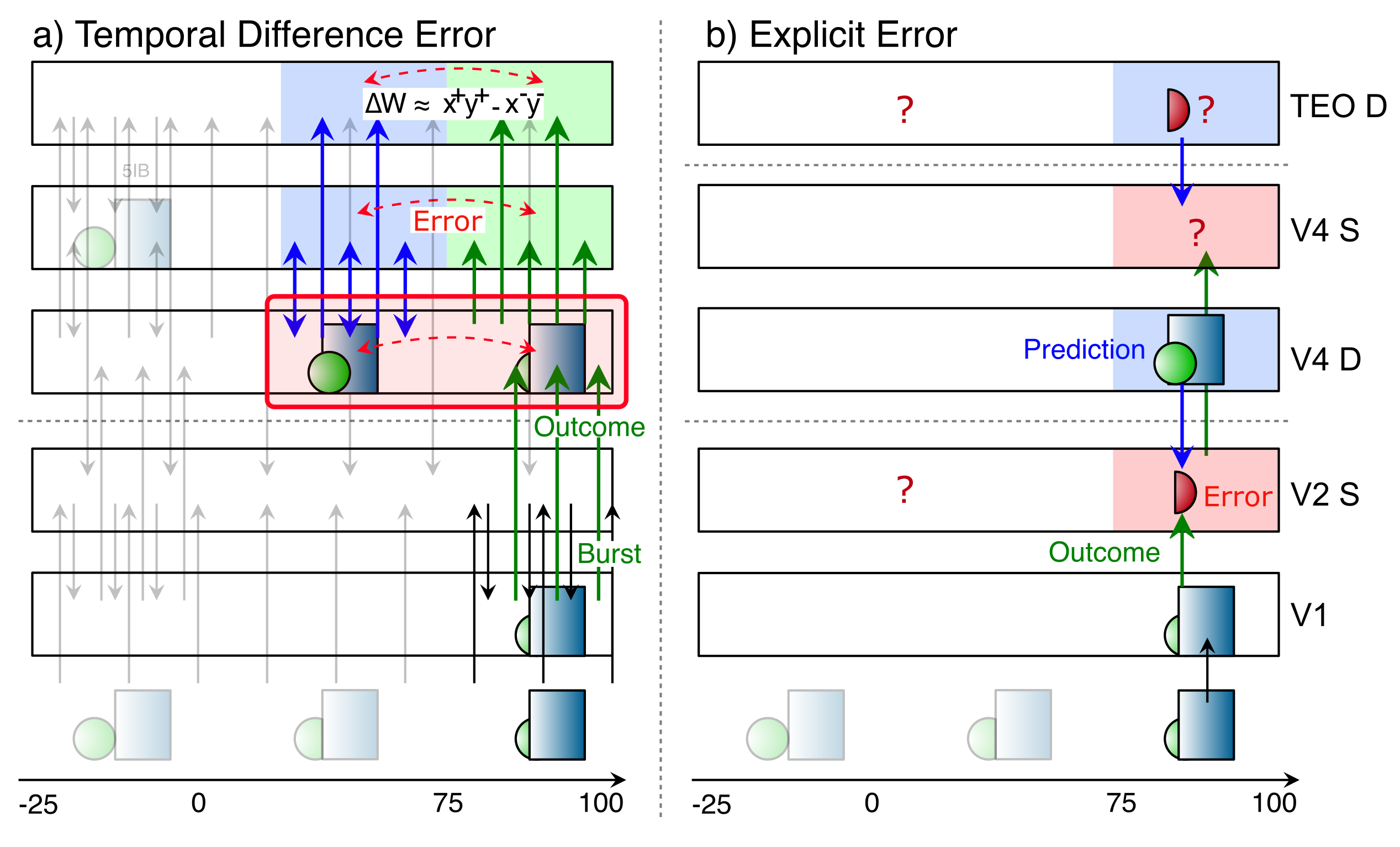}
  \caption{\footnotesize Comparison between: {\bf a)} The proposed thalamocortical temporal-difference predictive learning model (from Figure~\ref{fig.dltime}), versus {\bf b)} The Bayesian-style explicit error (EE) coding model (Rao \& Ballard, 1999; Friston, 2010, Bastos et al., 2012).  The EE model holds that superficial (S, lamina 2/3) error-coding neurons receive the prediction via a net inhibitory top-down projection from higher-level deep layer (D) neurons, and an excitatory bottom-up projection representing the outcome, such that their activation represents the difference.  To encode both signs of the error (omissions, false alarms) with positive-only spike rates, two separate populations of EE neurons would be required, or a more complicated deviation from tonic firing level scheme.  Unambiguous evidence of such EE coding neurons has not been found (Walsh et al, 2020).  In contrast, error signals in our proposed framework remain as a temporal difference between the two states of prediction vs. outcome, \emph{which enables all connectivity between cortical areas to be excitatory and always represent a positive encoding of either the prediction or outcome}.  In contrast, under EE, after one error subtraction at the lowest level, only error signals are hypothesized to flow forward to higher layers, meaning that the representations at higher layers are about increasingly higher-order \emph{errors}, not positive encodings of the environmental state at increasing levels of abstraction.  These are indicated by ? because they are difficult to picture intuitively, and they are inconsistent with extensive available data showing similar positive representations of the external world at all levels in the visual hierarchy.  Although some frameworks make claims about temporal dynamics, these are not strongly constrained by the basic computational framework, so that also remains a question.}
  \label{fig.ee}
\end{figure}

To further clarify the nature of the present theory, and introduce a body of relevant data, we contrast it with the widely-discussed explicit error (\emph{EE}) framework for predictive coding \citep{RaoBallard99,KawatoHayakawaInui93,Friston05,Friston10,OudenKokLange12,BastosUsreyAdamsEtAl12,LotterKreimanCox16} (Figure~\ref{fig.ee}).  The hypothesized locus for computing errors in this framework is in the superficial layers of the neocortex, which are suggested to directly compute the difference between bottom-up inputs from lower layers and top-down inputs from higher areas.  Despite many attempts to identify such explicit error-coding neurons in the cortex, no substantial body of unambiguous evidence has been discovered \citep{KokLange15,KokJeheedeLange12,SummerfieldEgner09,LeeMumford03,WalshMcGovernClarkEtAl20}.  Furthermore, due to the positive-only firing rate nature of neural coding, two separate populations would be required to convey both signs of prediction error signals, or it would have to be encoded as a variation from tonic firing levels, which are generally low in the neocortex.

By contrast, the use of temporal-difference error signals enables all connections between cortical layers to be excitatory and each layer can represent the positive encoding of either the prediction or outcome state, at different levels of abstraction.  These properties are overwhelmingly supported by extensive electrophysiological data about the hierarchical organization of representations, e.g., in the visual object recognition pathway \citep{KobatakeTanaka94,VanRullenThorpe02,CadieuHongYaminsEtAl14}, and are consistent with the widely-supported biased competition model for excitatory top-down attentional effects \citep{DesimoneDuncan95,ReynoldsChelazziDesimone99,MillerCohen01,OReillyWyatteHerdEtAl13}.

The EE approach requires net inhibitory top-down predictions, and it sends error signals forward, not positive representations of the actual state at a given level of abstraction.  Thus a literal interpretation (and at least one existing implementation; \citealp{LotterKreimanCox16}) has only error signals represented at all levels above the lowest level, which is inconsistent with the positive encoding of stimuli at various levels of abstraction across the visual hierarchy.  For example, although \citet{IssaCadieuDiCarlo18} observed an error-signal-like increase in activation for atypical faces in some pIT neurons, these neurons overall had a positive stimulus encoding, with only a relatively small, later, error-like modulation.

Furthermore, as discussed below, anticipatory predictions typically closely resemble the subsequent stimulus-driven activity, suggesting a positive, not inhibitory, effect  \citep{DuhamelColbyGoldberg92,LeeMumford03,CavanaghHuntAfrazEtAl10,WalshMcGovernClarkEtAl20}.  However, there are various different ways of reformulating the neural implementation of EE that can avoid some of these issues \citep{Spratling08,BastosUsreyAdamsEtAl12}, but perhaps this flexibility renders the framework difficult to falsify \citep{KogoTrengove15}.  In any case, an extensive treatment of the issues with EE is beyond the scope of this paper and has already been aptly covered by \citet{WalshMcGovernClarkEtAl20} --- our goal here is to highlight some of the core differences as a way to clarify the framework by way of contrast, and in relation to available data.

First, there are many examples of anticipatory predictive neural firing in the brain.  Of perhaps greatest relevance, \citet{BarczakOConnellMcGinnisEtAl18} recently showed that the auditory pulvinar in monkeys exhibits predictive firing using a carefully controlled auditory sequence that had no first-order acoustic differences from a background noise signal.  The pulvinar predictive activation preceded that of A1, suggesting a strong predictive role for pulvinar.  Unfortunately, the deep layers of higher auditory areas that should contribute to the formation of the pulvinar prediction were not recorded in this study, so their role in generating the prediction could not be determined.

Nevertheless, there is extensive additional evidence for top-down anticipatory activation of predicted stimuli, with activity patterns closely resembling the subsequent stimulus-driven ones \citep{WalshMcGovernClarkEtAl20}.  For example, the widely replicated predictive remapping effect, simulated in our model (Figure~\ref{fig.remap_units}) is of this nature \citep{DuhamelColbyGoldberg92,Wurtz08,CavanaghHuntAfrazEtAl10}.  The fact that these anticipatory activations are of a positive nature, consistent with the stimulus-driven activations, is inconsistent with the expected behavior of EE neurons, which should be inhibited by the top-down prediction, while not receiving any bottom-up stimulus.

However, the neural response to the actual predicted stimulus itself is typically suppressed relative to unexpected stimuli, i.e., \emph{expectation suppression} \citep{SummerfieldTrittschuhMontiEtAl08,TodorovicEdeMarisEtAl11,MeyerOlson11,BastosUsreyAdamsEtAl12}.  This phenomenon is widely cited as evidence in favor of the EE predictive coding framework, consistent with an inhibitory effect of the expectation.  Nevertheless, despite various conflicting results and many complications of interpretation, multiple comprehensive reviews conclude that it is difficult to distinguish expectation suppression from the neural adaptation effects that underlie the well-documented \emph{repetition suppression} effect \citep{WalshMcGovernClarkEtAl20,VinkenVogels17,KokLange15,KokJeheedeLange12,SummerfieldEgner09,LeeMumford03}.  Furthermore, detailed single-neuron level recordings are the least likely to show these effects --- instead, they are most evident in aggregate signals such as the BOLD response in fMRI, suggesting that they may more strongly reflect population-level differences in activity, rather than individual explicit error coding neurons.

As noted earlier, accurately predicted outcomes in our framework would result in a continued adaptation of the neural response carrying over from the prediction to the outcome state, whereas unexpected outcomes would be associated with two distinct patterns of activity over a given area: first the prediction and then the outcome.  Thus, the unexpected outcome state would not be subject to the prior neural adaptation effects, and furthermore the time-integrated aggregate activity over these two patterns would be greater compared to the single activity state associated with an accurately predicted outcome.  Thus, our model explains expectation suppression without invoking EE neurons, meaning that considerably more detailed and replicable experimental paradigms using single-neuron resolution techniques are needed to distinguish EE from our framework.

\subsection{Alpha Frequency Effects}

The alpha frequency bursting of 5IB neurons acting as drivers into the pulvinar naturally entrains the predictive learning process in our model to this fundamental rhythm, which has long been recognized as an important signature of posterior cortical function \citep{Berger29,Walter53,NunnOsselton74,VarelaToroJohnEtAl81,VanRullenKoch03}.  A number of different functional associations with alpha have been established, and this literature is large and growing rapidly.  Thus, we refer the reader to recent reviews \citep{JensenBonnefondMarshallEtAl15,VanRullen16,ClaytonYeungKadosh18,FosterAwh19} while highlighting the data most relevant to our specific framework here, organized according to a set of key points.

\begin{itemize}
	\item \emph{Alpha is specifically associated with deep neocortical layers and the pulvinar, and with feedback pathways in the cortex.}  This has been established using direct laminar-specific electrophysiological single-neuron and local field potential (LFP) recordings \citep{LuczakBarthoHarris13,BuffaloFriesLandmanEtAl11,MaierAdamsAuraEtAl10,MaierAuraLeopold11,SpaakBonnefondMaierEtAl12,XingYehBurnsEtAl12}, and feedforward vs. feedback manipulations \citep{vonSteinChiangKonig00,vanKerkoerleSelfDagninoEtAl14,BastosVezoliBosmanEtAl15,JensenBonnefondMarshallEtAl15,MichalareasVezolivanPeltEtAl16}.  	These data are consistent with the 5IB alpha bursting and the major role of cortical deep layers in driving top-down corticocortical projections (in addition to the 6CT pathway which is specific to the pulvinar).  By contrast, these same papers show that superficial cortical layers are associated with gamma frequency (40 Hz) dynamics.  However, the next point raises some important interpretational difficulties.

	\item \emph{Increases in cortical activity levels, e.g., due to attention, produce a corresponding decrease in alpha power, while decreased activity increases alpha power}  \citep{WordenFoxeWangEtAl00,KellyLalorReillyEtAl06,KlimeschSausengHanslmayr07,FriesWomelsdorfOostenveldEtAl08,JensenMazaheri10,FosterAwh19}. This pattern is not exactly what you might expect if alpha was a signature of predictive learning.  However, given that these same pulvinar and thalamocortical pathways are also widely regarded as important for attention \citep{LaBergeBuchsbaum90,BenderYouakim01,SnowAllenRafalEtAl09,SaalmannKastner11,ZhouSchaferDesimone16,FiebelkornKastner19}, this pattern presents a challenge for many theorists.  However, it is possible to explain this pattern as arising directly from the desynchronizing effects of cortical activity on alpha power.  Specifically, neural spiking is associated with broadband noise, due to the highly random, Poisson nature of spike firing, which can desynchronize the entrainment of lower-frequency oscillations including alpha \citep{WaldertLemonKraskov13,RayMaunsell11,PrivmanMalachYeshurun13,SolomonKragelSperlingEtAl17}.  In other words, because cortical activity is inherently noisy, it tends to interfere with the coherent activity across populations of neurons needed to produce a strong alpha frequency power signal.  This explanation is directly supported by studies manipulating and measuring cortical activity \citep{ZhouSchaferDesimone16,FriesWomelsdorfOostenveldEtAl08}, and is consistent with alpha power changes being a \emph{result} of attentional modulation, but not their cause \citep{AntonovChakravarthiAndersen20}.  Thus, while attention and predictive learning can both affect overall activity levels in cortex, and thus drive changes in alpha power, alpha power itself is not a transparent measure of the underlying mechanisms supporting these functions, which may help to explain some contradictory patterns of results \citep{FosterAwh19,GundlachMorattiForschackEtAl20,KeitelKeitelBenwellEtAl19}.

	\item \emph{Alpha phase effects provide a more direct measure of thalamocortical function than alpha power, and have been more consistently related to perception, attention, and prediction}  \citep{NunnOsselton74,VarelaToroJohnEtAl81,VanRullenKoch03,BuschDuboisVanRullen09,MathewsonFabianiGrattonEtAl10,PalvaPalva11,JaegleRo13,NeupaneGuittonPack17,Solis-VivancoJensenBonnefond18}.  For example, weak, near-threshold stimuli are more reliably detected and processed when presented in the trough of the individual's ongoing alpha cycle.  Of greatest relevance to the present paper are studies showing effects of prediction on alpha phase \citep{SamahaBauerCimaroliEtAl15,MayerSchwiedrzikWibralEtAl16,ShermanKanaiSethEtAl16}.  For example, \citet{MayerSchwiedrzikWibralEtAl16} showed that prestimulus alpha phase directly correlated with the predictability of the upcoming stimulus, and the pattern of this prestimulus activation was indistinguishable from the subsequent stimulus activation pattern.  This is consistent with our model, and less consistent with the EE framework, as discussed previously.  \citet{NeupaneGuittonPack17} found strong alpha coherence effects in LFP recordings distributed across V4, associated with the predictive remapping of receptive fields \citep{DuhamelColbyGoldberg92}.

	\item \emph{Discrete, salient, or oscillatory stimuli entrain the alpha cycle in the brain} \citep{SpaakLangeJensen14,MathewsonPrudhommeFabianiEtAl12}.  Furthermore, the massive literature on \emph{event related potentials} (ERPs) may represent a significant contribution from alpha-level entrainment \citep{MakeigWesterfieldJungEtAl02,GruberKlimeschSausengEtAl05,Klimesch11}.  These entrainment effects are consistent with the 5IB entrainment mechanisms in our framework, as described earlier, and entrainment is functionally important for aligning predictive learning with relevant salient or unexpected outcomes.

	\item \emph{The pulvinar contributes to synchronizing alpha phase relationships across different brain areas} \citep{SaalmannPinskWangEtAl12,FiebelkornPinskKastner18}.  This is consistent with the broad, convergent pattern of projections into the pulvinar from many different cortical areas, and the corresponding broad projections back out to these same areas \citep{Shipp03,ArcaroPinskKastner15}.  Functionally, this convergence and synchronization is important for integrating the contributions from these different areas at the same time, to generate predictions over the pulvinar.

	\item \emph{The theta cycle, comprised of a pair of alpha cycles, organizes saccades, and attentional, motor, and mnemonic processes} \citep{FiebelkornKastner19}.  The theta rhythm is dominant in the medial temporal lobe and hippocampus, and has been extensively studied there \citep{KahanaSeeligMadsen01,Buzsaki05}.  Furthermore, there is a  sharp peak of saccade fixation durations at 200 ms, which suggests that two alpha cycles are typically required for  complete processing of a given fixation.  On the first cycle, the predictions from before the eye moved may be fairly vague depending on factors such as the size of the saccade and familiarity with the environment.  But after the first alpha cycle of a fixation, a subsequent \emph{postdiction} phase can provide an important additional learning opportunity, to consolidate and more deeply encode the current fixation (computationally equivalent to an auto-encoder).  Also, a mix of smaller saccades (including microsaccades) and larger saccades enables a range of more and less predictable outcomes on the first alpha cycle after the saccade, and matches human behavior \citep{Martinez-CondeOtero-MillanMacknik13,Martinez-CondeMacknikHubel04}.

\end{itemize}

Putting all of these points together, a particularly effective way of testing the predictions of our framework would be measuring alpha phase changes emerging in the prestimulus period as a function of predictive learning in predictable sequential stimulus streams.  In addition, it would also be important to examine theta and alpha-cycle dynamics in relation to predictive learning in the context of attention, motor control, and memory processes, to better understand the larger systems-level temporal organization of learning and processing in the brain \citep{FiebelkornKastner19}.

\subsection{Predictions for Predictive Learning}

In this section, we enumerate a set of direct, testable predictions from our framework.  Before doing so, there are several important considerations for any experimental test of the theory.  First, the nature of what is to be learned must be matched to the pulvinar area in question.  For example, learning a new variation of basic physics in movies at the alpha time scale (e.g., altering properties such as gravity, inertia, or elasticity), would be appropriate for the lower level visual pathways.  At higher visual levels (e.g., IT cortex), it might be possible to use simple sequences of different objects, although it is not clear to what extent the hippocampus or prefrontal cortex might also contribute in this case \citep{GavornikBear14,FiserMahringerOyiboEtAl16}.  To distinguish pulvinar learning effects from pervasive motor learning supported by other brain areas, it would be most effective to directly measure activity in the pulvinar and / or associated perceptual neocortical areas, instead of involving overt behavioral performance.

Much of the learning in posterior sensory cortex should take place early in development, requiring very early developmental interventions or genetic knockouts that are expressed from the start (which can also have other interpretational issues if not highly selective).  In our models, the bulk of the basic sensory predictive learning happens very quickly, because the basic first-level regularities are quite strong and relatively easily learned.  While there are longer-term changes in the higher-level pathways in our models, more fine-grained measurements would likely be required to see these changes.  Once this learning has taken place, the remaining contributions of the thalamocortical circuit are likely more strongly weighted toward its role in attention, as we discuss below.  Finally, directly lesioning or inactivating the pulvinar is not likely to be very informative, because existing work has shown dramatic effects on cortical activity \citep{ZhouSchaferDesimone16,PurushothamanMarionLiEtAl12}, and also any effects could be attributed to the attentional contributions of the pulvinar.

With these considerations in mind, here are a set of strong predictions from our model that should be testable using existing techniques.  Failure to obtain the predicted result, while adhering to all the relevant constraints, would constitute a falsification of our model.

\begin{itemize}
	\item \emph{Blocking 5IB bursting mechanisms early in developmental learning should disrupt learning}.  It should be possible to selectively knock out or modify the channels that cause this specific population of neurons to burst fire, and doing so should have a significant effect on learning in associated neocortical and pulvinar areas, given the critical role that this burst firing plays on the predictive learning process as elaborated above.

	\item \emph{Blocking synaptic plasticity in pulvinar (specifically the 6CT inputs) very early in developmental learning should impair learning}.  While most of the learning overall should occur in the neocortex as a result of the temporal difference error signal broadcast by the pulvinar (which should remain generally intact), learning in the 6CT projections is important, especially right at the start, to map the emerging neocortical representations into the space defined by the 5IB projections.

	\item \emph{Temporal differences on an alpha cycle timescale actually drive synaptic plasticity in an error-driven learning manner, in neocortical pyramidal neurons and in 6CT inputs to pulvinar}.  That is, if a pre / post pair of neurons across a synapse is more active in the prediction than the subsequent outcome, the synapse should experience LTD (long term depression), and vice-versa if the activity pattern is reversed (long term potentiation, LTP, for more activity in outcome than prediction).  Furthermore, if activity is essentially stable across both prediction and outcome phases, then weights should not change (modulo a small level of Hebbian learning; \citealp{OReillyMunakata00,OReillyMunakataFrankEtAl12}).  This should be directly testable using current  experimental methods, and is perhaps the single most important empirical test of this entire framework, and it also underlies many other current approaches to error-driven learning in the brain \citep{BengioMesnardFischerEtAl17,WhittingtonBogacz19,LillicrapSantoroMarrisEtAl20}.  One general consideration is the extent to which an awake \emph{in vivo} preparation would be required to capture all the neuromodulatory and other factors present when this learning normally takes place.  Some suggestive evidence in such a preparation is generally consistent with a sensitivity to relatively short-term temporal dynamics \citep{LimMcKeeWoloszynEtAl15}, although these results lacked the direct measurement of individual neural activity across a synapse.

\end{itemize}

\section{Discussion}

We have hypothesized a novel computational function for the distinctive features of thalamocortical circuits \citep{ShermanGuillery06,UsreySherman18}, as supporting a specific form of prediction-error driven learning, where predictions arise from the numerous top-down layer 6CT projections into the pulvinar, and the strong, sparse, focal driving 5IB inputs supply the bottom-up sensory-driven outcome. The phasic bursting nature of the 5IB inputs results in a natural temporal-difference error signal of prediction followed by outcome, consistent with extensive neural recording data.  This temporal dynamic is also essential for enabling predictions to be generated without contamination from current sensory inputs, and predicts a characteristic alpha frequency prediction cycle based on the 10hz bursting cycle of the 5IB inputs, consistent with the pervasive influence of alpha on perception and neural dynamics \citep{JensenBonnefondMarshallEtAl15,VanRullen16,ClaytonYeungKadosh18,FosterAwh19}.  In short, the hypothesized predictive learning function fits remarkably well with a number of well-established properties of these thalamocortical circuits, and we also provided a set of additional predictions that could be tested to further evaluate this theory, especially in contrast to the widely-discussed alternative of explicit error coding neurons, which have not been unambiguously supported across a range of empirical studies \citep{WalshMcGovernClarkEtAl20}.

Furthermore, we implemented this theory in a large scale model of the visual system, and demonstrated that learning based strictly on predicting what will be seen next is, in conjunction with a number of critical biologically motivated network properties and mechanisms, capable of generating abstract, invariant categorical representations of the overall shapes of objects.  The nature of these shape representations closely matches human shape similarity judgments on the same objects.  Thus, predictive learning has the potential to go beyond the surface structure of its inputs, and develop systematic, abstract encodings of the environment.   We found that comparison models based on standard error backpropagation learning did not learn a categorical structure that went beyond the surface similarity present in the visual input layers, and future work is focused on narrowing down the specific mechanisms required to drive this learning.

In addition to the predictive learning functions of the deep / thalamic layers, these same circuits are also likely critical for supporting powerful top-down attentional mechanisms that have a net multiplicative effect on superficial-layer activations \citep{BortoneOlsenScanziani14,OlsenBortoneAdesnikEtAl12,BortoneOlsenScanziani14,OlsenBortoneAdesnikEtAl12}. The importance of the pulvinar for attentional processing has been widely documented \cite[e.g.,]{LaBergeBuchsbaum90,BenderYouakim01,SaalmannPinskWangEtAl12}, and there is likely an additional important role of the thalamic reticular nucleus (TRN), which can contribute a surround-inhibition contrast-enhancing effect on top of the incoming attentional signal from the cortex \citep{Crick84,Pinault04,WimmerSchmittDavidsonEtAl15,JaramilloMejiasWang19}. In other work in progress, we have shown that the deep / thalamic circuits in our model produce attentional effects consistent with the abstract \citet{ReynoldsHeeger09} model, while the contributions of the deep layer networks to this function are broadly consistent with the folded-feedback model \citep{Grossberg99}.  These attentional modulation signals cause the bidirectional constraint satisfaction process in the superficial network to focus on task-relevant information while down-regulating responses to irrelevant information --- in the real world, there are typically too many objects to track at any given time, so predictive learning must be directed toward the most important objects \citep{Pylyshyn89,CavanaghHuntAfrazEtAl10,RichterdeLange19}.

There is also data suggesting that the pulvinar is important for supporting \emph{confidence} judgments, driven by relative ambiguity in a random dot motion categorization task \citep{KomuraNikkuniHirashimaEtAl13}.  Critically for the present framework, this confidence modulation only emerged in the period after the first 100 ms of processing, and manifested as a positive correlation with confidence (i.e., more unambiguous stimuli resulted in higher firing rates).  We can interpret this as reflecting an ongoing generative \emph{postdiction} of the stimulus signal, with stronger firing associated with more unambiguous top-down activation based on the current internal representation.  Note that this directionality is the opposite of explicit error-coding neurons, which would presumably increase with increasing error / ambiguity in the prediction.  Interestingly, inactivation of these pulvinar neurons resulted in a substantial (200\%) increase in opt-out choices on the most ambiguous stimuli, suggesting a level of metacognitive awareness of the pulvinar signal (or at least a direct effect of pulvinar on relevant metacognitive processes).  Predictive accuracy would be an ideal source of metacognitive confidence signals across a wide range of domains, suggesting another important contribution of pulvinar even after initial learning.  \citet{JaramilloMejiasWang19} present a comprehensive model of attentional, decision-making, and working memory contributions of the pulvinar, including this confidence data, which is generally compatible with our framework, although it does not address any learning phenomena.

Considerable further work remains to be done to more precisely characterize the essential properties of our biologically motivated model necessary to produce this abstract form of learning, and to further explore the full scope of predictive learning across different domains.  We strongly suspect that extensive cross-modal predictive learning in real-world environments, including between sensory and motor systems, is a significant factor in infant development and could greatly multiply the opportunities for the formation of higher-order abstract representations that more compactly and systematically capture the structure of the world \citep{YuSmith12}.  Future versions of these models could thus potentially provide novel insights into the fundamental question of how deep an understanding a pre-verbal human, or a non-verbal primate, can develop \citep{SpelkeBreinlingerMacomberEtAl92,ElmanBatesKarmiloff-SmithEtAl96}, based on predictive learning mechanisms.  This would then represent the foundation upon which language and cultural learning builds, to shape the full extent of human intelligence.

\clearpage


\section{Appendix}

\input{deep_pred_lrn_2021_supp}

\clearpage

\bibliography{ccnlab}

\end{document}

%% file: deep_pred_lrn_2021_supp.tex
All of the materials described here, including the experimental study, the computational models, and the code to perform the representational similarity analysis, are all available on our github account at: \url{https://github.com/ccnlab/deep-obj-cat} and the new version of the {\em emergent} simulation environment is at: \url{https://github.com/emer/leabra} which contains extensive documentation and examples that can be run in Python or the Go language.  The best place to start in understanding computationally how the predictive learning model works is with the FSA model described in the main text, which is available at: \url{https://github.com/emer/leabra/tree/master/examples/deep_fsa}.  For the large and complex WWI model, the most complete understanding can only be had by directly examining the code, as there are a number of details that are not efficiently captured in this Appendix text.

\section{Representational Similarity Analysis Methods}

The different representations being compared here are:
\begin{description}
\item[Leabra:] The DeepLeabra (biological model) TE layer representations (specifically TEs = superficial -- results are very similar for deep as well).
\item[Bp:] The TEs layer representations from the backpropagation version of biological model, including {\em What, Where} and {\em What * Where} integration layers, trained with the V1p and V1hp (low and high resolution pulvinar) layers as final output layers, using the time $t$ target pattern from the $t-1$ input (i.e., as a predictive network).
\item[V1:] The gabor-filtered representation of the visual input to both of the above models, which was identical across them.
\item[PredNet:] Highest layer (6th Layer) of the PredNet architecture.

\item[Expt:] Similarity matrix constructed from human pairwise similarity judgments (see {\em Behavioral Experiment Methods}).
\end{description}

\begin{figure}
  \begin{tabular}{llll}
   \multicolumn{2}{c}{Centroid} & \multicolumn{2}{c}{Bp} \\
	 \parbox[t]{1.4in}{\raggedright  	\setstretch{.5} \small
	\begin{enumerate}
	\item pyramid
	\begin{itemize}[leftmargin=*]
	\item banana
	\item layercake
	\item trafficcone
	\item sailboat
	\item trex
	\end{itemize}
	\item vertical
	\begin{itemize}[leftmargin=*]
	\item person
	\item guitar
	\item tablelamp
	\end{itemize}
	\item round
	\begin{itemize}[leftmargin=*]
	\item doorknob
	\item donut
	\end{itemize}
	\end{enumerate}
	} & 
	 \parbox[t]{1.4in}{\raggedright 	\setstretch{.5} \small
	\begin{enumerate}
	\item[3.] round cont'd
	\begin{itemize}[leftmargin=*]
	\item handgun
	\item chair
	\end{itemize}
	\item[4.] box
	\begin{itemize}[leftmargin=*]
	\item slrcamera
	\item elephant
	\item piano
	\item fish
	\end{itemize}
	\item[5.] horiz
	\begin{itemize}[leftmargin=*]
	\item car
	\item heavycannon
	\item stapler
	\item motorcycle
	\end{itemize}
	\end{enumerate}
	} &
	 \parbox[t]{1.4in}{\raggedright  	\setstretch{.5} \small
	\begin{enumerate}
	\item cat1
	\begin{itemize}[leftmargin=*]
	\item banana
	\item layercake
	\item trafficcone
	\item sailboat
	\item trex
	\item person
	\item guitar
	\item tablelamp
	\item doorknob
	\item donut
	\end{itemize}
	\end{enumerate}
	} & 
	 \parbox[t]{1.4in}{\raggedright 	\setstretch{.5} \small
	\begin{enumerate}
	\item[1.] cat1 cont'd
	\begin{itemize}[leftmargin=*]
	\item handgun
	\item chair
	\item slrcamera
	\item elephant
	\item piano
	\item fish
	\item car
	\end{itemize}
	\item[2.] cat2
	\begin{itemize}[leftmargin=*]
	\item heavycannon
	\item stapler
	\item motorcycle
	\end{itemize}
	\end{enumerate}
	}\\
   \multicolumn{2}{c}{V1} &
   \multicolumn{2}{c}{PredNet} \\
	 \parbox[t]{1.4in}{\raggedright  	\setstretch{.5} \small
	\begin{enumerate}
	\item cat1
	\begin{itemize}[leftmargin=*]
	\item trafficcone
	\item sailboat
	\item person
	\item guitar
	\item tablelamp
	\item chair
	\end{itemize}
	\item cat2
	\begin{itemize}[leftmargin=*]
	\item layercake
	\item trex
	\item doorknob
	\item donut
	\end{itemize}
	\end{enumerate}
	} & 
	 \parbox[t]{1.4in}{\raggedright 	\setstretch{.5} \small
	\begin{enumerate}
	\item[2.] cat2 cont'd
	\begin{itemize}[leftmargin=*]
	\item handgun
	\item slrcamera
	\item elephant
	\item piano
	\item fish
	\item car
	\item heavycannon
	\item stapler
	\item motorcycle
	\end{itemize}
	\item[3.] cat3
	\begin{itemize}[leftmargin=*]
	\item banana
	\end{itemize}
	\end{enumerate}
	} & 
	 \parbox[t]{1.4in}{\raggedright  	\setstretch{.5} \small
	\begin{enumerate}
	\item cat1
	\begin{itemize}[leftmargin=*]
	\item trafficcone
	\item sailboat
	\item person
	\item guitar
	\item tablelamp
	\item layercake
	\end{itemize}
	\item cat2
	\begin{itemize}[leftmargin=*]
	\item trex
	\item donut
	\item banana
	\item handgun
	\end{itemize}
	\end{enumerate}
	} & 
	 \parbox[t]{1.4in}{\raggedright 	\setstretch{.5} \small
	\begin{enumerate}
	\item[2.] cat2 cont'd
	\begin{itemize}[leftmargin=*]
	\item slrcamera
	\item elephant
	\item fish
	\item car
	\item heavycannon
	\item stapler
	\item motorcycle
	\end{itemize}
	\item[3.] cat3
	\begin{itemize}[leftmargin=*]
	\item chair
	\item doorknob
	\item piano
	\end{itemize}
	\end{enumerate}
	}\\
	\end{tabular}
	\caption{\footnotesize Shape categories used for similarity matrix plots in main paper.  {\em Centroid} shape categories are near-best for both the Leabra model and the Expt results, and fit our visual intuitions about overall shape. {\em Bp} are reliably optimal for Bp model from all starting points.  {\em V1} are reliably optimal for V1 inputs, and also were close to the best for the Bp and PredNet layer 6 representations.  {\em PredNet} are best stable solution for PredNet layer 6.}
	\label{fig.cats}
\end{figure}

An optimal category cluster can be defined as one that has high within-cluster similarity and low between-cluster similarity.  This can be operationalized by the {\em contrast} distance metric, based on a 1-correlation ({\em dissimilarity}) measure, as the difference between the average within-cluster similarity and the average between-cluster similarity:
\begin{equation}
 cd = \langle 1-r_{in} \rangle - \langle 1-r_{out} \rangle 
\end{equation}
With distance-like 1-correlation values, this contrast distance should be minimized (it is typically negative), or equivalently the contrast on raw correlation values can be maximized (it is typically a positive number -- just the sign flip of distance value).  We refer to the positive numbers and maximization here as that is more intuitive.

Starting with an initial set of clusters, a permutation-based hill-climbing strategy was used to determine a local minimum in this measure: each item was tested in each of the other possible categories, and if that configuration reduced the overall average contrast distance metric across all items, then it was adopted and the process iterated until no such permutation improved the metric.  This algorithm can only decrease the number of clusters (by moving all items out of a given cluster), so different numbers of initial clusters can be used to search the overall space.

Figure~\ref{fig.cats} shows the resulting categories. The Bp model converged on the same cluster state from all starting configurations tested, varying from 5 to 2 initial categories.  This is the cluster set shown in Figure~\ref{fig.bpred} of the main paper, and has an average contrast distance ({\em acd}) of 0.0838 (this is relatively low because the patterns were overall quite similar).  Likewise, the V1 patterns (which were the same across Leabra and Bp models) reliably converged on the same pattern (shown in Figure~\ref{fig.bpred}), with {\em acd} = 0.2448.

For the PredNet layer 6 representations, starting from the V1 categories gave the best results of any other set ({\em acd} = 0.1967), and a few permutations resulted in a reliable solution that was arrived at from all other 3 category starting points tested, shown in Figure~\ref{fig.cats} ({\em acd} = 0.2820).  This indicates that PredNet did not go much beyond the structure present in the input, even though it did not use the V1 gabor filtering used in the Leabra and Bp models (i.e., this V1-level encoding well-captures the structure of the visual inputs in general).  The PredNet pixel and layer 1 representations both converged on essentially a single monolithic category with very low acd (0.0018, 0.0013).

For the Leabra TE representations, we found a set of {\em centroid} shape categories that are near-best when considering both the Leabra model and the results from the human behavioral experiment (Expt).  Starting from these categories, the permutation analysis converged on reducing the size of the vertical and round categories to one item each, over a sequence of 5 steps.  This is consistent with the observation from Figure~\ref{fig.rsa} that there are three broader categories within which the 5 finer-grained categories are embedded (i.e., vertical and pyramid are overall similar to each other, as are round and box).  Nevertheless, our initial visual intuition about the broad shape categories, along with a bias against having single-item categories, reinforced the use of the finer-grained centroid selection.  The average contrast difference of our centroid selection is 0.5071, while the maximal result from the permutation was 0.5526, which is a relatively small proportional difference.

Furthermore, once we had collected the human experimental data ({\em Expt}), it was clear that it strongly coincided with our original shape intuitions, and with the finer-grained 5 category centroid structure.  Starting from the centroid categories, the maximal permutation made only 3 changes, moving trex (T-rex) and handgun into the horizontal category, and chair into the pyramid, going from a distance score of 0.3083 to 0.3225, which is a relatively small improvement.  However, using the maximal {\em Expt} clusters directly on the Leabra model gives a lower {\em acd} measure of 0.3745 (compared to 0.5071 for centroid), so the centroid categories represent a good middle-ground between experiment and the model, and this strong shared similarity structure with near-optimal cluster structures confirms that the model and people are encoding largely the same information.

In contrast, if we organize the experiment similarity matrix using the Bp categories, it produces a very poor average contrast distance measure of 0.0643 (compared to 0.3083 for the centroid categories), strongly suggesting that people's shape representations are not compatible with that simple structure.

\begin{figure}
  \centering\includegraphics[width=6in]{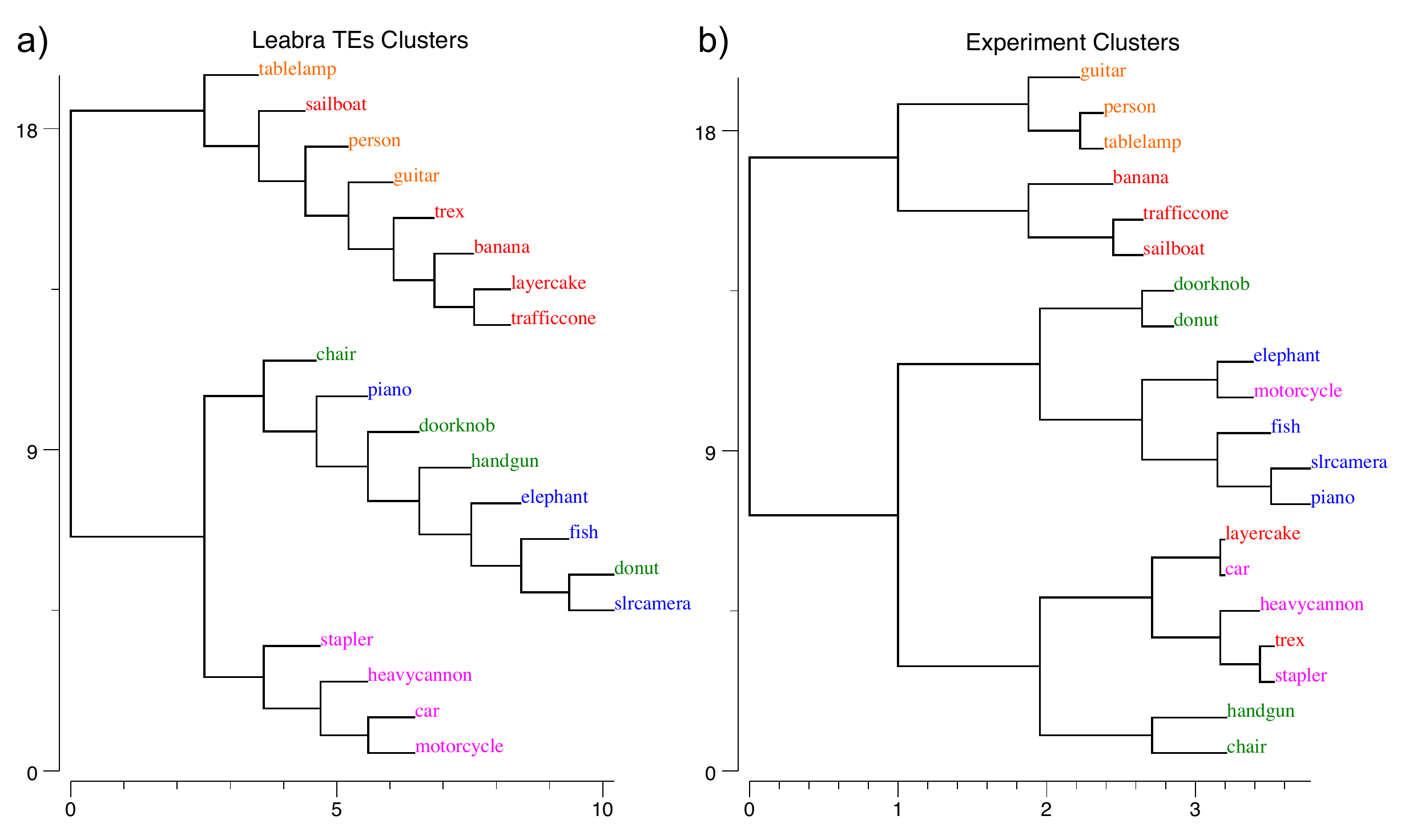}
  \caption{\footnotesize  Agglomerative clustering on the Leabra and Expt representations, with the centroid categories color coded.  The most reliable information from this is the leaf-level groupings, as the rest of the structure is indeterminante and history dependent in reducing higher-dimensional structure down to a 2D plot.  Both cluster plots show a strong tendency to group leaf items together in the same centroid categories, with a few exceptions in each case.  Also, the Leabra plot nicely captures the broader 3-category structure evident in the similarity matrix plots, within which the 5 finer-grained centroid categories are organized.  Overall, this provides further confirmation that the model and the human subjects are organizing the shapes in largely the same way.}
  \label{fig.clust}
\end{figure}

Another approach to determining clusters from similarity matrices, {\em agglomerative clustering}, starts with all items as singletons, and iteratively combines the closest two into a new cluster.  The results for the Leabra and Expt similarity matrices are shown in Figure~\ref{fig.clust}, which has also color-coded the items in terms of their category status according to the centroid structure.  Due to a strong history dependency in the clustering process, and the indeterminacy of reducing a high-dimensional similarity structure down to two dimensions, structure beyond the leaf level is not very reliable (ties are also broken by a random number generator), but nevertheless you can clearly see that in both cases items from the same cluster are almost always together as leaves in the plots.  This then provides additional converging support for the idea that the model is learning the same kind of shape categories as people have.

For the network layer RSA computations, activation vectors were accumulated separately for each 3D object item, and within that separately for each frame index of the movie.  To be able to monitor similarity metrics as the model trained, we used a running-average integration of neural activity across trials to accumulate the patterns.  Specifically, the current activation pattern across each layer was recorded and averaged unit-by-unit with a time constant of $\tau = 10$. Critically, by integrating separately for each frame, this running-average computation did not introduce any bias for temporally-adjacent frames to be more similar.  Nevertheless, when we computed the frame-to-frame similarities for TE, they were quite high (.901 correlation on average across all objects).

\section{Behavioral Experiment Methods}

\begin{figure}
  \centering\includegraphics[width=4in]{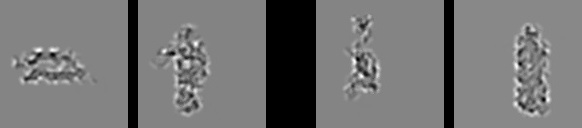}
  \caption{\footnotesize Example stimulus from the behavioral experiment, using the V1 reconstruction of the actual input images presented to the model, to better capture the coarse-grained perception of the model.  Subjects were requested to choose which of the two pairs, Left or Right, was most similar in terms of {\em overall shape}.}
  \label{fig.expt}
\end{figure}

The behavioral experiment was conducted on Amazon.com's MTurk web platform under University of Colorado IRB approval (19-0176), using 30 participants each categorizing up to 800 image pairs as shown in Figure~\ref{fig.expt}, using the standard {\em simple image categorization} framework with a lightly customized script.  Objects were drawn from the 156 3D object set, but data was aggregated in terms of the 20 basic-level categories (car, stapler, etc) because we could not sample all 156 x 156 object pairs.  Thus, the resulting data was aggregated for each category pair in terms of the proportion of times when that pair was selected when presented.

The individual images were produced by reconstructing from the V1 transform that the computational model used in its high resolution V1 input layer, to give human participants as similar of an experience as possible to how the model ``saw'' the objects, and to reduce the influence of existing semantic knowledge which was entirely missing in our model (Figure~\ref{fig.expt}).

\section{Biological Model Methods}

This section provides more information about the {\em DeepLeabra} {\em What-Where Integration (WWI)} model.  The purpose of this information is to give more detailed insight into the model's function beyond the level provided in the main text, but with a model of this complexity, the only way to really understand it is to explore the model itself.  It is available for download at: \url{https://github.com/ccnlab/deep-obj-cat/sims/cemer}.  We now have a full replication of this model in our new, much more transparent simulation framework, available at \url{https://github.com/ccnlab/deep-obj-cat/sims/wwi3d} --- this is more readable and recommended.  Furthermore, the best way to understand this model is to understand the framework in which it is implemented, which is explained in great detail, with many running simulations explaining specific elements of functionality, at \url{http://CompCogNeuro.org}

\subsection{Layer Sizes and Structure}

\begin{table}
  \centering
\begin{tabular}{llrrlll}
\hline
     &      & \multicolumn{2}{c}{{\bf Units}} & \multicolumn{2}{c}{{\bf Pools}} & \\
{\bf Area} & {\bf Name} & {\bf X} & {\bf Y} & {\bf X} & {\bf Y} & {\bf Receiving Projections} \\
\hline
V1 & V1s & 4 & 5 & 8 & 8 &  \\
   & V1p & 4 & 5 & 8 & 8 & V1s V2d V3d V4d TEOd  \\
V1h & V1hs & 4 & 5 & 16 & 16 &  \\
   & V1hp & 4 & 5 & 16 & 16 & V1s V2d V3d V4d TEOd  \\
Eyes & EyePos & 21 & 21 & & &  \\
     & SaccadePlan & 11 & 11 & & &  \\
     & Saccade & 11 & 11 & & &  \\
Obj & ObjVel & 11 & 11 & & & \\
V2 & V2s & 10 & 10 & 8 & 8 & V1s LIPs V3s V4s TEOd V1p V1hp \\
   & V2d & 10 & 10 & 8 & 8 & {\bf V2s} V1p V1hp LIPd LIPp V3d V4d V3s TEOs \\
LIP & MtPos& 1 & 1 & 8 & 8 & V1s \\
    & LIPs & 4 & 4 & 8 & 8 & MtPos ObjVel SaccadePlan EyePos LIPp \\
    & LIPd & 4 & 4 & 8 & 8 & {\bf LIPs} LIPp ObjVel Saccade EyePos \\
    & LIPp & 1 & 1 & 8 & 8 & {\bf MtPos} V1s LIPd \\
V3 & V3s & 10 & 10 & 4 & 4 & V2s V4s TEOs DPs LIPs V1p V1hp DPp TEOd \\
   & V3d & 10 & 10 & 4 & 4 & {\bf V3s} V1p V1hp DPp LIPd DPd V4d V4s DPs TEOs \\
   & V3p & 10 & 10 & 4 & 4 & {\bf V3s} V2d DPd TEOd \\
DP & DPs & 10 & 10 & & & V2s V3s TEOs V1p V1hp V3p TEOp \\
   & DPd & 10 & 10 & & & {\bf DPs} V1p V1hp DPp TEOd \\
   & DPp & 10 & 10 & & & {\bf DPs} V2d V3d DPd TEOd \\
V4 & V4s & 10 & 10 & 4 & 4 & V2s TEOs V1p V1hp \\
   & V4d & 10 & 10 & 4 & 4 & {\bf V4s} V1p V1hp V4p TEOd TEOs \\
   & V4p & 10 & 10 & 4 & 4 & {\bf V4s} V2d V3d V4d TEOd \\
TEO & TEOs & 10 & 10 & 4 & 4 & V4s V1p V1hp TEs\\
   & TEOd & 10 & 10 & 4 & 4 & {\bf TEOs TEOd} V1p V1hp V4p TEOp TEp TEd \\
   & TEOp & 10 & 10 & 4 & 4 & {\bf TEOs} V3d V4d TEOd TEd \\
TE & TEs & 10 & 10 & 4 & 4 & TEOs V1p V1hp \\
   & TEd & 10 & 10 & 4 & 4 & {\bf TEs TEd} V1p V1hp V4p TEOp TEp TEOd \\
   & TEp & 10 & 10 & 4 & 4 & {\bf TEs} V3d V4d TEOd \\
\hline
\end{tabular}
\caption{\footnotesize Layer sizes, showing numbers of units in one pool (or entire layer if Pool is missing), and the number of Pools of such units, along X,Y axes.  Each area has three associated layers: {\em s} = superficial layer, {\em d} = deep layer (context updated by 51B neurons in same area, shown in bold), {\em p} = pulvinar layer (driven by 5IB neurons from associated area, shown in bold).}
\label{tab.layer_sizes}
\end{table}

Figure~\ref{fig.model} in the main text shows the general configuration of the model, and Table~\ref{tab.layer_sizes} shows the specific sizes of each of the layers, and where they receive inputs from. 

All the activation and general learning parameters in the model are at their standard Leabra defaults.

\subsection{Projections}

\begin{figure}
  \centering\includegraphics[width=5in]{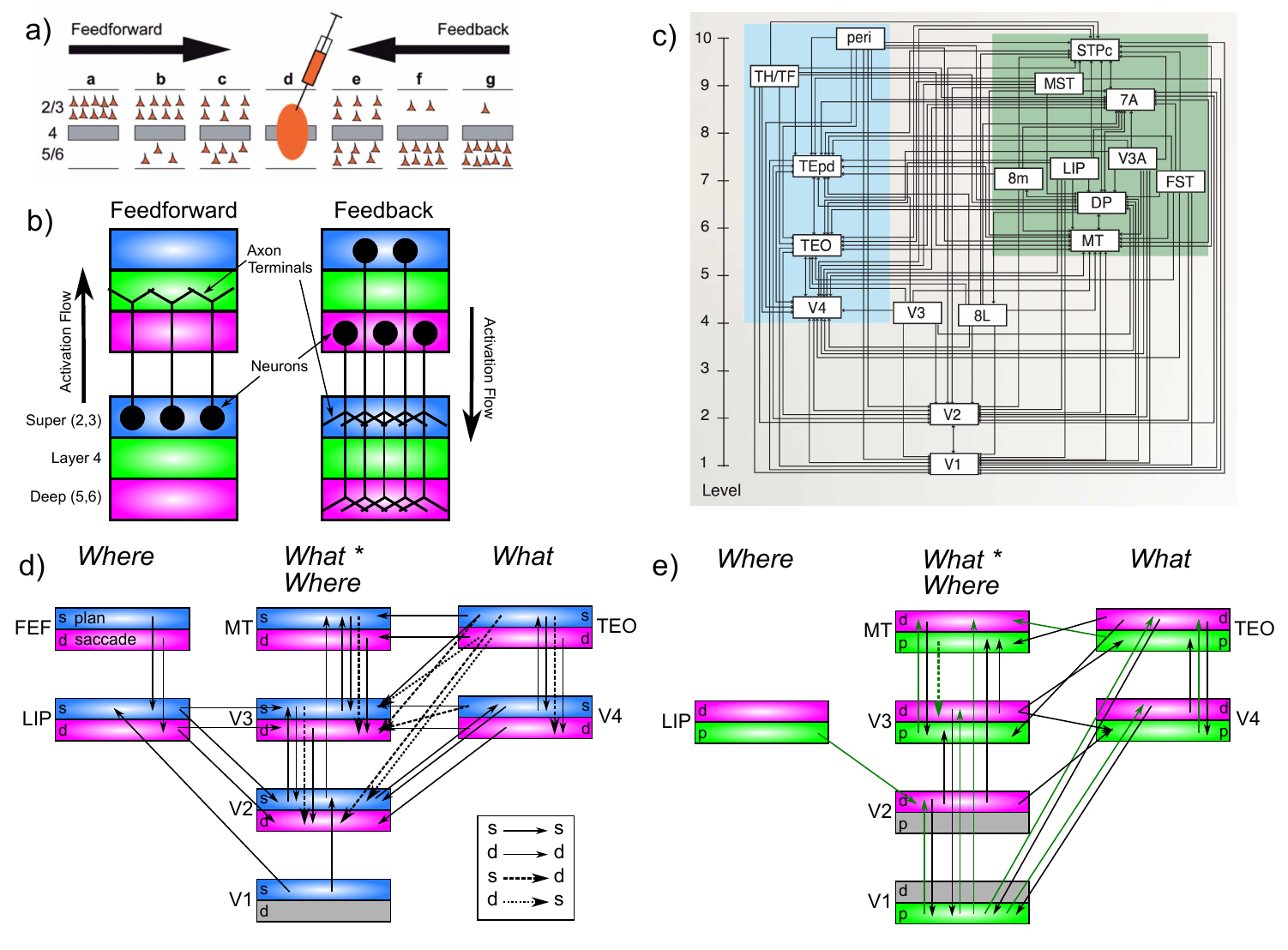}
  \caption{\footnotesize Principles of connectivity in DeepLeabra.  {\bf a)} Markov et al (2014) data showing density of {\em retrograde} labeling from a given injection in a middle-level area (d): most feedforward projections originate from superficial layers of lower areas (a,b,c) and deep layers predominantly contribute to feedback (and more strongly for longer-range feedback).  {\bf b)} Summary diagram showing most feedforward connections originating in superficial layers of lower area, and terminating in layer 4 of higher area, while feedback connections can originate in either superficial or deep layers, and in both cases terminate in both superficial and deep layers of the lower area (adapted from Felleman \& Van Essen, 1991). {\bf c)} Anatomical hierarchy as determined by percentage of superficial layer source labeling (SLN) by Markov et al (2014) --- the hierarchical levels are well matched for our model, but we functionally divide the dorsal pathway (shown in green background) into the two separable components of a {\em Where} and a {\em What * Where} integration pathway.  {\bf d)} Superficial and deep-layer connectivity in the model.  Note the repeating motif between hierarchically-adjacent areas, with bidirectional connectivity between superficial layers, and feedback into deep layers from both higher-level superficial and deep layers, according to canonical pattern shown in panels a and b.  Special patterns of connectivity from TEO to V3 and V2, involving crossed super-to-deep and deep-to-super pathways, provide top-down support for predictions based on high-level object representations. {\bf e)} Connectivity for deep layers and pulvinar in the model, which generally mirror the corticocortical pathways (in d).  Each pulvinar layer (p) receives 5IB driving inputs from the labeled layer (e.g., V1p receives 5IB drivers from V1).  In reality these neurons are more distributed throughout the pulvinar, but it is computationally convenient to organize them together as shown.  Deep layers (d) provide predictive input into pulvinar, and pulvinar projections send error signals (via temporal differences between predictions and actual state) to {\em both} deep and superficial layers of given areas (only d shown).  Most areas send deep-layer prediction inputs into the main V1p prediction layer, and receive reciprocal error signals therefrom.  The strongest constraint we found was that pulvinar outputs (colored green) must generally project only to higher areas, not to lower areas, with the exceptions of DPp $\rightarrow$ V3 and LIPp $\rightarrow$ V2.  V2p was omitted because it is largely redundant with V1p in this simple model.}
  \label{fig.model_cons}
\end{figure}

The general principles and patterns of connectivity are shown in Figure~\ref{fig.model_cons} (and Figures~\ref{fig.sg06} and \ref{fig.dltime} in the main text).  As noted in the main text, the connectivity and overall structure obeys the established principles identified in neocortical anatomy \citep{RocklandPandya79,FellemanVanEssen91,MarkovVezoliChameauEtAl14,MarkovErcsey-RavaszGomesEtAl14}.

Detailing each of the specific parameters associated with the different projections shown in Table~\ref{tab.layer_sizes} would take too much space --- those interested in this level of detail should download the model from the link shown above.  There are topographic projections between many of the lower-level retinotopically-mapped layers, consistent with our earlier vision models \citep{OReillyWyatteHerdEtAl13}.  For example the 8x8 unit groups in V2 are reduced down to the 4x4 groups in V3 via a 4x4 unit-group topographic projection, where neighboring units have half-overlapping receptive fields (i.e., the field moves over 2 unit groups in V2 for every 1 unit group in V3), and the full space is uniformly tiled by using a wrap-around effect at the edges.  Similar patterns of connectivity are used in standard deep convolutional neural networks.  However, we do not share weights across units as in a true convolutional network.

The projections from ObjVel (object velocity) and SaccadePlan layers to LIPs, LIPd were initialized with a topographic sigmoidal pattern that moved as a function of the position of the unit group, by a factor of .5, while the projections from EyePos were initialized with a gaussian pattern.  These patterns multiplied uniformly distributed random weights in the .25 to .75 range, with the lowest values in the topographic pattern having a multiplier of .6, while the highest had a multiplier of 1 (i.e., a fairly subtle effect).  This produced faster convergence of the LIP layer when doing {\em Where} pathway pre-training compared to purely random initial weights, consistent with \citet{PougetSejnowski97a} and related work on parietal gain field basis function representations.

In addition to exploring different patterns of overall connectivity, we also explored differences in the relative strengths of receiving projections, which can be set with a \texttt{wt\_scale.rel} parameter in the simulator.  All feedforward pathways have a default strength of 1.  For the feedback projections, which are typically weaker (consistent with the biology), we explored a discrete range of strengths, typically .5, .2, .1, and .05.  The strongest top-down projections were into V2s from LIP and V3, while most others were .2 or .1.  Likewise projections from the pulvinar were weaker, typically .1.  These differences in strength sometimes had large effects on performance during the initial bootstrapping of the overall model structure, but in the final model they are typically not very consequential for any individual projection.

\subsection{Training Parameters}

\begin{figure}
  \centering\includegraphics[width=5.5in]{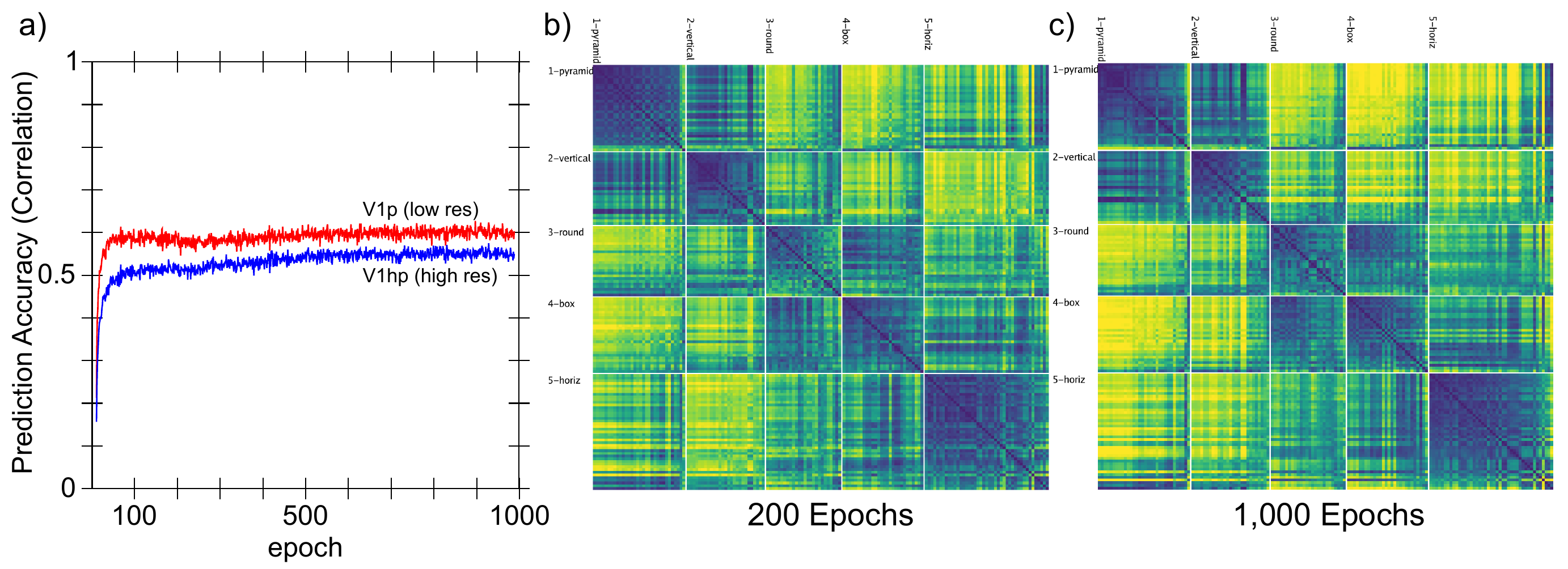}
  \caption{\footnotesize {\bf a)} Predictive learning curve for DeepLeabra, showing the correlation between prediction and actual over the two different V1 layers.  Initial learning is quite rapid, followed by a slower but progressive learning process that reflects development of the IT representations (e.g., manipulations that interfere with those areas selectively impair this part of the learning curve).  Overall prediction accuracy remains far from perfect, as shown in Figure~\ref{fig.model} in main text, and significantly worse than the backpropagtion-based models.  This is a typical finding from Leabra models which are significantly more constrained as a result of bidirectional attractor dynamics, Hebbian learning, and inhibitory competition -- i.e., the very things that are likely important for forming abstract catgorical representations. {\bf b)} Similarity matrix over TEs layer at 200 epochs, which has less contrast and definition (particularly evident in the off-block-diagonal differences) compared to the 1,000 epoch result ({\bf c} also shown in Figure~\ref{fig.rsa} in main text).}
  \label{fig.lrn}
\end{figure}

Training typically consisted of 512 alpha trials per epoch (51.2 seconds of real time equivalent), for 1,000 such epochs. Each trial was generated from a virtual reality environment in the emergent simulator, that rendered first-person views with moving eye position onto the object tumbling through space with fixed motion and rotation parameters over the sequence of 8 frames (see Figure~\ref{fig.model} in main text for representative example).  Each frame was rendered at 256 x 256 resolution, and processed through our standard V1 gabor filters which are described in detail in \citet{OReillyWyatteHerdEtAl13}.

Because the start of each sequence of 8 frames is unpredictable, we turned off learning for that trial, which improves learning overall.  We have recently developed an automatic such mechanism based on the running-average (and running variance) of the prediction error, where we turn off learning whenever the current prediction error z-normalized by these running average values is below 1.5 standard deviations, which works well, and will be incorporated into future models.  Biologically, this could correspond to a connection between pulvinar and neuromodulatory areas that could regulate the effective learning rate in this way.

Figure~\ref{fig.lrn}a shows the learning trajectory of the model, indicating that it learns quite rapidly.  This rapid initial learning is likely facilitated by the extensive use of shortcut connections convering from all over the simulated visual system onto the V1 pulvinar layers, and direct projections back from these pulvinar layers.  Thus, error signals are directly communicated and can drive learning quickly and efficiently.  However, there are also extensive indirect, bidirectional connections among the superficial layers, which can drive indirect error backpropagation learning as well.

\subsection{Model Algorithms}

The biologically-based model was implemented using the Leabra framework, which is described in detail in previous publications \citep{OReillyHazyHerd16,OReillyMunakataFrankEtAl12,OReillyMunakata00,OReilly98,OReilly96}, and summarized here.  The online textbook at \url{https://CompCogNeuro.org} provides the most comprehensive description of the framework, while \url{https://github.com/emer/leabra} has a summary of all the equations (and the code itself).  There are two main implementations of Leabra, one in the {\em C++ emergent} software, and a new one using Go and Python language at the prior link.  These same equations and standard parameters have been used to simulate over 40 different models in \citet{OReillyMunakataFrankEtAl12,OReillyMunakata00}, and a number of other research models.  Thus, the model can be viewed as an instantiation of a systematic modeling framework using standardized mechanisms, instead of constructing new mechanisms for each model \citep{OReillyHazyHerd16}.

The neurons use a rate code version of the adaptive exponential (AdEx) conductance-based point neuron model \citep{BretteGerstner05}, with the standard RC circuit equations:
\begin{equation}
 \Delta V_m(t) = \tau \sum_c g_c(t) \overline{g_c} (E_c - V_m(t)),
 \label{eq.vm}
\end{equation}
where $c$ represents excitatory, inhibitory, and leak channels.  Inhibition is driven by simulated interneurons in proportion to feedforward and feedback dynamics, producing sparse distributed representations, and controlling the effects of bidirectional excitatory connections between layers.

Each neuron learns using a more biologically-based version of the Contrastive Hebbian Learning (CHL) algorithm, as shown in Figure~\ref{fig.dltime}:
\begin{equation}
 \Delta_{chl} = x^+ y^+ - x^- y^-
 \label{eq:chl}
\end{equation}
where $x$ is the sending activation, $y$ is the receiving activation, and the $+$ superscript indicates activations in the plus phase, and $-$ those in the minus phase.  The actual learning equations, detailed at \url{https://github.com/emer/leabra} and in the online textbook: \url{https://CompCogNeuro.org} produce a combination of error-driven and self-organizing factors, which  emerge out of a single learning rule that was derived from a biologically detailed model of synaptic plasticity by \citep{UrakuboHondaFroemkeEtAl08}, and is closely related to the Bienenstock, Cooper \& Munro (BCM) algorithm \citep{BienenstockCooperMunro82}.

\subsubsection{Deep Context}

This section describes in detail the equations that are specific to the {\em Deep} version of Leabra that implements the specific predictive learning additions to the general algorithm.  Like the simple recurrent network (SRN) \citep{Elman90,Jordan89} which the deep predictive learning model functionally resembles, the primary computational specialization required is the maintenance of prior temporal context in the CT layer.  In addition, the pulvinar layers have to be driven by the bottom-up inputs in the plus phase, after being driven by the CT inputs in the minus phase.

Computationally, the CT layer is specialized for maintaining context from the previous alpha cycle, to generate the prediction over the pulvinar layer.  At the end of every plus phase, a new CT context excitatory input is computed from the normalized dot product of the context weights times the sending activations, just as in the standard net input used in Leabra:
\begin{equation}
  \eta_j = \langle x_i w_{ij} \rangle = \oneo{n} \sum_i x_i w_{ij}
 \label{eq.netin_ti}
\end{equation}
where $x_i$ are the sending activations and $w_{ij}$ are the weights.  This net input is then added in with the standard net input at each cycle of processing during the subsequent alpha cycle.

The relative strength of these context layer inputs was set progressively larger for higher layers in the network, with a maximum of 4 in V4, TEO, and TE.  In addition, TEO and TE received \emph{self} context projections which provide an extended window of temporal context into the prior 200 ms interval, consistent with multiple sources of neural data \citep{ChaudhuriKnoblauchGarielEtAl15}.  These self projections were connected only within the narrower Pool level of units, enabling these neurons to develop mutually-excitatory loops to sustain activations over the multiple trials when the same object was present.  We hypothesize that these modifications correspond to biological adaptations in IT cortex that likewise support greater sustained activation of object-level representations.

Learning of the context weights occurs as normal, but using the sending activation states from the {\em prior} time step's activation.

\subsubsection{Computational and Biological Details of SRN-like Functionality}

Predictive auto-encoder learning has been explored in various frameworks, but the most relevant to our model comes from the application of the SRN to a range of predictive learning domains \citep{Elman90,ElmanBatesKarmiloff-SmithEtAl96}.  One of the most powerful features of the SRN is that it enables error-driven learning, instead of arbitrary parameter settings, to determine how prior information is integrated with new information.  Thus, SRNs can learn to hold onto some important information for a relatively long interval, while rapidly updating other information that is only relevant for a shorter duration.  This same flexibility is present in our DeepLeabra model.  Furthermore, because this temporal context information is hypothesized to be present in the deep layers throughout the entire neocortex (in every microcolumn of tissue), the DeepLeabra model provides a more pervasive and interconnected form of temporal integration compared to the SRN, which typically just has a single temporal context layer associated with the internal ``hidden'' layer of processing units.

An extensive computational analysis of what makes the SRN work as well as it does, and explorations of a range of possible alternative frameworks, has led us to an important general principle: {\em subsequent outcomes determine what is relevant from the past}.  At some level, this may seem obvious, but it has significant implications for predictive learning mechanisms based on temporal context.  It means that the information encoded in a temporal context representation cannot be learned at the time when that information is presently active.  Instead, the relevant contextual information is learned on the basis of what happens next.

\begin{figure}
  \centering\includegraphics[width=3in]{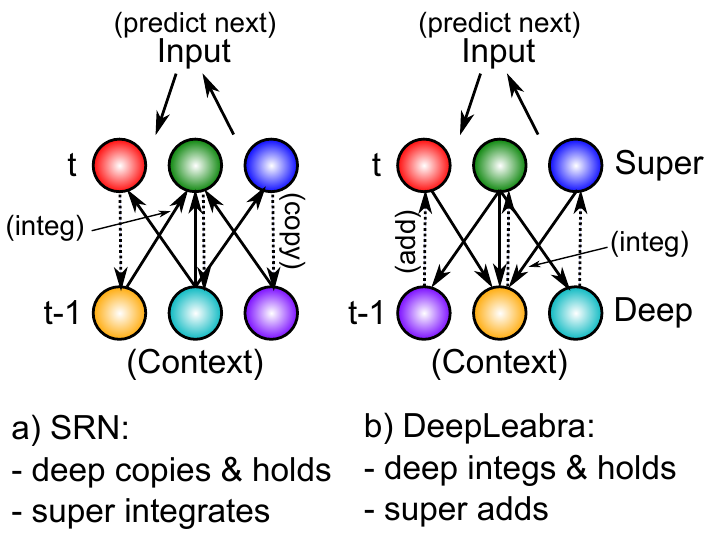}
  \caption{\footnotesize  How the DeepLeabra temporal context computation compares to the SRN mathematically. {\bf a)} In a standard SRN, the context (deep layer biologically) is a copy of the hidden activations from the prior time step, and these are held constant while the hidden layer (superficial) units integrate the context through learned synaptic weights.  {\bf b)} In DeepLeabra, the deep layer performs the weighted integration of the soon-to-be context information from the superficial layer, and then holds this integrated value, and feeds it back as an additive net-input like signal to the superficial layer.  The context net input is pre-computed, instead of having to compute this same value over and over again.  This is more efficient, and more compatible with the diffuse interconnections among the deep layer neurons.  Layer 6 projections to the thalamus and back recirculate this pre-computed net input value into the superficial layers (via layer 4), and back into itself to support maintenance of the held value.}
  \label{fig.srn_vs_ti}
\end{figure}

This explains the peculiar power of the otherwise strange property of the SRN: the temporal context information is preserved as a {\em direct copy} of the state of the hidden layer units on the previous time step (Figure~\ref{fig.srn_vs_ti}), and then learned synaptic weights integrate that copied context information into the next hidden state (which is then copied to the context again, and so on).  This enables the error-driven learning taking place in the {\em current} time step to determine how context information from the {\em previous} time step is integrated.  And the simple direct copy operation eschews any attempt to shape this temporal context itself, instead relying on the learning pressure that shapes the hidden layer representations to also shape the context representations.  In other words, this copy operation is essential, because there is no other viable source of learning signals to shape the nature of the context representation itself (because these learning signals require future outcomes, which are by definition only available later).

The direct copy operation of the SRN is however seemingly problematic from a biological perspective: how could neurons copy activations from another set of neurons at some discrete point in time, and then hold onto those copied values for a duration of 100 ms, which is a reasonably long period of time in neural terms (e.g., a rapidly firing cortical neuron fires at around 100 Hz, meaning that it will fire 10 times within that context frame).  However, there is an important transformation of the SRN context computation, which is more biologically plausible, and compatible with the structure of the deep network (Figure~\ref{fig.srn_vs_ti}). Specifically, instead of copying an entire set of activation states, the context activations (generated by the phasic 5IB burst) are immediately sent through the adaptive synaptic weights that integrate this information, which we think occurs in the 6CC (corticortical) and other lateral integrative connections from 5IB neurons into the rest of the deep network.

The result is a {\em pre-computed net input} from the context onto a given hidden unit (in the original SRN terminology), not the raw context information itself.  Computationally, and metabolically, this is a much more efficient mechanism, because the context is, by definition, unchanging over the 100 ms alpha cycle, and thus it makes more sense to pre-compute the synaptic integration, rather than repeatedly re-computing this same synaptic integration over and over again (in the original feedforward backpropagation-based SRN model, this issue did not arise because a single step of activation updating took place for each context update --- whereas in our bidirectional model many activation update steps must take place per context update).

There are a couple of remaining challenges for this transformation of the SRN.  First, the pre-computed net input from the context must somehow persist over the subsequent 100 ms period of the alpha cycle.  We hypothesize that this can occur via NMDA and mGluR channels that can easily produce sustained excitatory currents over this time frame.  Furthermore, the reciprocal excitatory connectivity from 6CT to TRC and back to 6CT could help to sustain the initial temporal context signal.  Second, these contextual integration synapses require a different form of learning algorithm that uses the sending activation from the prior 100 ms, which is well within the time constants in the relevant calcium and second messenger pathways involved in synaptic plasticity.

\section{Backpropagation Model Methods}

\begin{figure}
  \centering\includegraphics[width=2.5in]{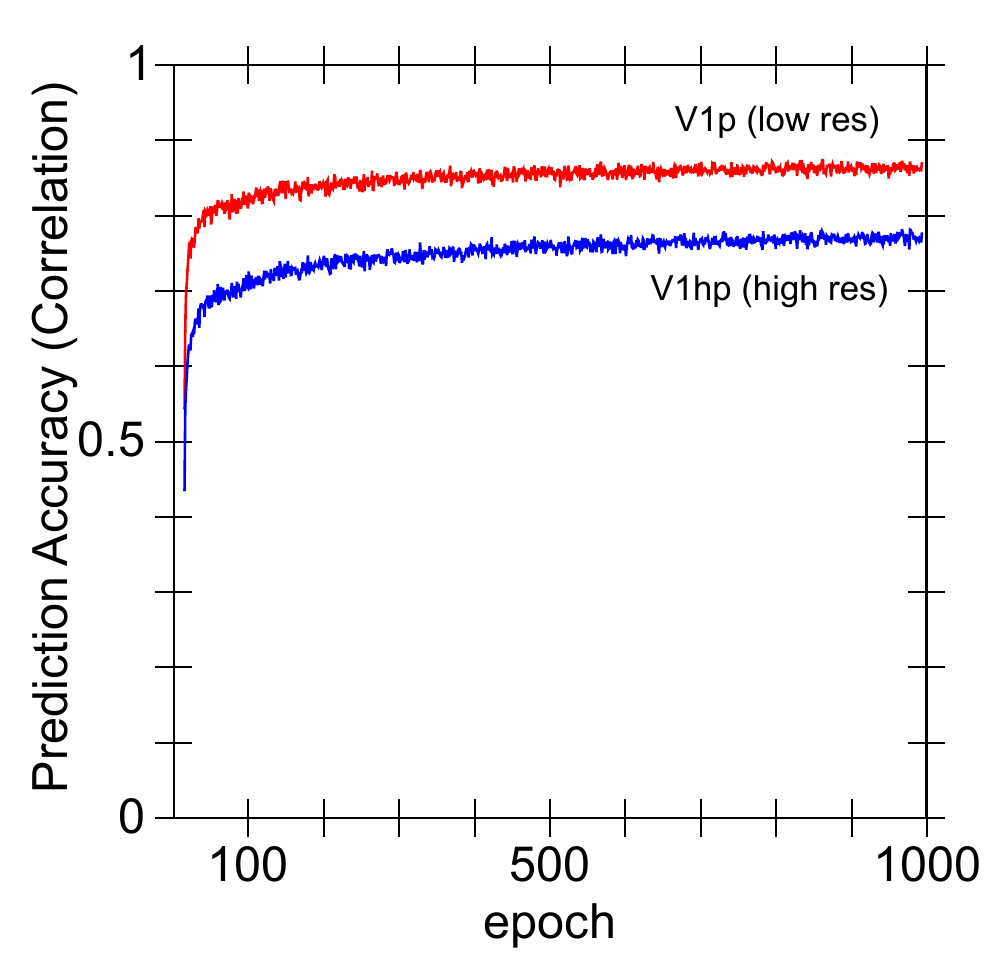}
  \caption{\footnotesize Learning curves for the backpropagation version of the WWI model.  Although it achieves better predictive accuracy than the DeepLeabra version, it fails to acquire abstract object category structure, indicating a potential tradeoff between simplifying and categorizing inputs, versus predicting precisely where the low-level visual features will move.}
  \label{fig.bp_lrn}
\end{figure}

The backpropagation version of the WWI model has exactly the same layer sizes and {\em feedforward} patterns of connectivity as the DeepLeabra version.  Topographically, the V1p and V1hp pulvinar layers serve as output layers at the highest level of the network, receiving all the various connections from deep layers as shown in Table~\ref{tab.layer_sizes}.  Likewise, the LIPp served as a target output layer for the Where pathway.  To achieve predictive learning, the V1 pulvinar targets were from the scene at time $t$, while the V1s inputs were from the scene at time $t-1$.  We also ran a comparison auto-encoder model that had inputs and target outputs from the same time step, and it showed even less systematic organization of its higher-level representations, further supporting the notion that predictive learning is important, across all frameworks.  The learning curve for the predictive version is shown in Figure~\ref{fig.bp_lrn}, which shows better overall prediction accuracy compared to the DeepLeabra model.  However, as the RSA showed, this backpropagation model failed to learn object categories that go beyond the input similarity structure, indicating that perhaps it was paying too much ``attention'' in learning to this low-level structure, and lacked the necessary mechanisms to enable it to impose a simplifying higher-level structure on top of these inputs.

\section{PredNet Model Methods}

The PredNet architecture was designed to incorporate principles from predictive coding theory into a neural network model for predicting the next frame in a video sequence. Details of the model can be found in the original paper \citep{LotterKreimanCox16}, but here we provide a brief overview of the architecture. 

\subsection{Architecture}

PredNet is a deep convolutional neural network that is composed of layers containing discrete modules. The lowest layer generates a prediction of incoming inputs (i.e. the pixels in the next frame), while each of the higher layers attempts to predict the {\em errors} made by the previous layer. Each layer contains an input convolutional module ($A_l$), a recurrent representational module ($R_l$), a prediction module ($\hat{A}_l$), and a representation of its own errors ($E_l$). The input convolutional module ($A_l$) transforms its input with a set of standard convolutional filters, a rectified linear activation function, and a max-pooling operation. The recurrent representation module ($R_l$) is a convolutional LSTM, which is a recurrent convolutional network that replaces the matrix multiplications in the standard LSTM equations with convolutions, allowing it to maintain a spatially organized representation of its inputs over time. The prediction module ($\hat{A_l}$) consists of another standard convolutional layer and rectified linear activation that is used to generate predictions from the output of $R_l$. These predictions are then compared against the output of the input convolutional module ($A_l$). The errors generated in this comparison are represented explicitly in $E_l$, which applies a rectified linear activation to a concatenation of the positive ($A_l - \hat{A}_l$) and negative ($\hat{A}_l - A_l$) prediction errors. These errors then become the inputs to the next layer. 

\begin{equation}
A_l^t = 
\begin{cases}
    x_t, & \text{if } l = 0\\
    MaxPool(ReLU(Conv(E_{l-1}^t))), & \text{if } l > 0
\end{cases}
\end{equation}
\begin{equation}
\hat{A}_l^t = ReLU(Conv(R_l^t))
\end{equation}
\begin{equation}
E_l^t = [ReLU(A_l^t - \hat{A}_l^t); ReLU(\hat{A}_l^t - A_l^t)]
\end{equation}
\begin{equation}
R_l^t = ConvLSTM(E_l^{t-1},R_l^{t-1},UpSample(R_{l+1}^t))
\end{equation}

At each time step in the video sequence, PredNet generates a prediction of the next frame. This is done as follows: first, the $R_l$ is computed for each layer starting from the top of the hierarchy (because each $R_l^t$ depends on input from $R_{l+1}^t$), and then the $A_l^t$, $\hat{A}_l^{t}$ and $E_l^t$ are computed in a feed-forward fashion (becauase each $A_l^t$ depends on input from the layer below, $E_{l-1}^t$). 


All analyses in the RSA were conducted using the representations from the $R_l$ layers. 

\begin{figure}
  \centering\includegraphics[width=2.5in]{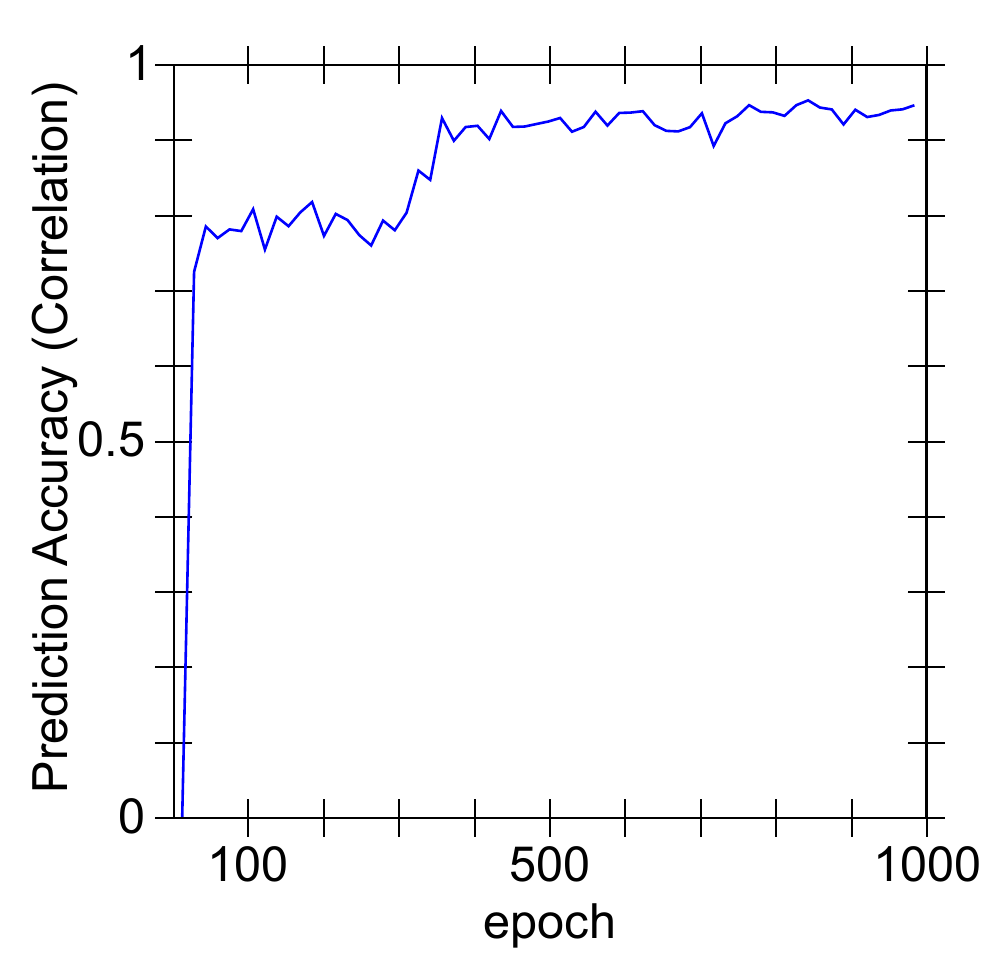}
  \caption{\footnotesize Learning curves for the PredNet model.  This model achieves the best overall prediction performance but also has the least well differentiated, categorical representations.}
  \label{fig.prednet_lrn}
\end{figure}

\subsection{Implementation details}

All experiments with the PredNet architecture were performed using PyTorch. An informal hyperparameter search was conducted to find the settings that maximized representational similarity to the human judgments. This was done by conducting RSA on each layer for each hyperparameter setting, and computing, according to the Centroid categories derived from the human data, the difference between the average within-category similarity and the average between-category similarity. Our final architecture had 6 layers with 3, 16, 32, 64, 128, and 256 filters in the $A_l$ and $R_l$ modules, and 3x3 kernels throughout the whole network. We also found that using sigmoid and tanh activation functions in fully-connected convolutional LSTMs slightly improved performance, so these were used for all experiments.

The weights in the PredNet model are trained using error backpropagation. Predictions are generated and errors are computed at all levels of the hierarchy, but the model performs better when only the lowest layer's errors are backpropagated \citep{LotterKreimanCox16}. We confirmed these results with experiments that backpropagated the errors in higher layers, in which performance (in terms of mean squared error) was marginally reduced but the RSA results were similar. For this reason, all reported experiments used a PredNet that was trained by only backpropagating the lowest level error.

The model was trained using a batch size of 8 and an Adam optimizer with a learning rate of 0.0001, with no scheduler, for 150,000 batches.  A training curve is shown in Figure~\ref{fig.prednet_lrn}, showing that it achieves the best overall prediction accuracy of any model we tested, and yet does not have representations that are as differentiated or categorical as our biologically based model, as shown in the main paper.

\subsection{Regularization experiments}

\begin{figure}
  \centering\includegraphics[width=3.5in]{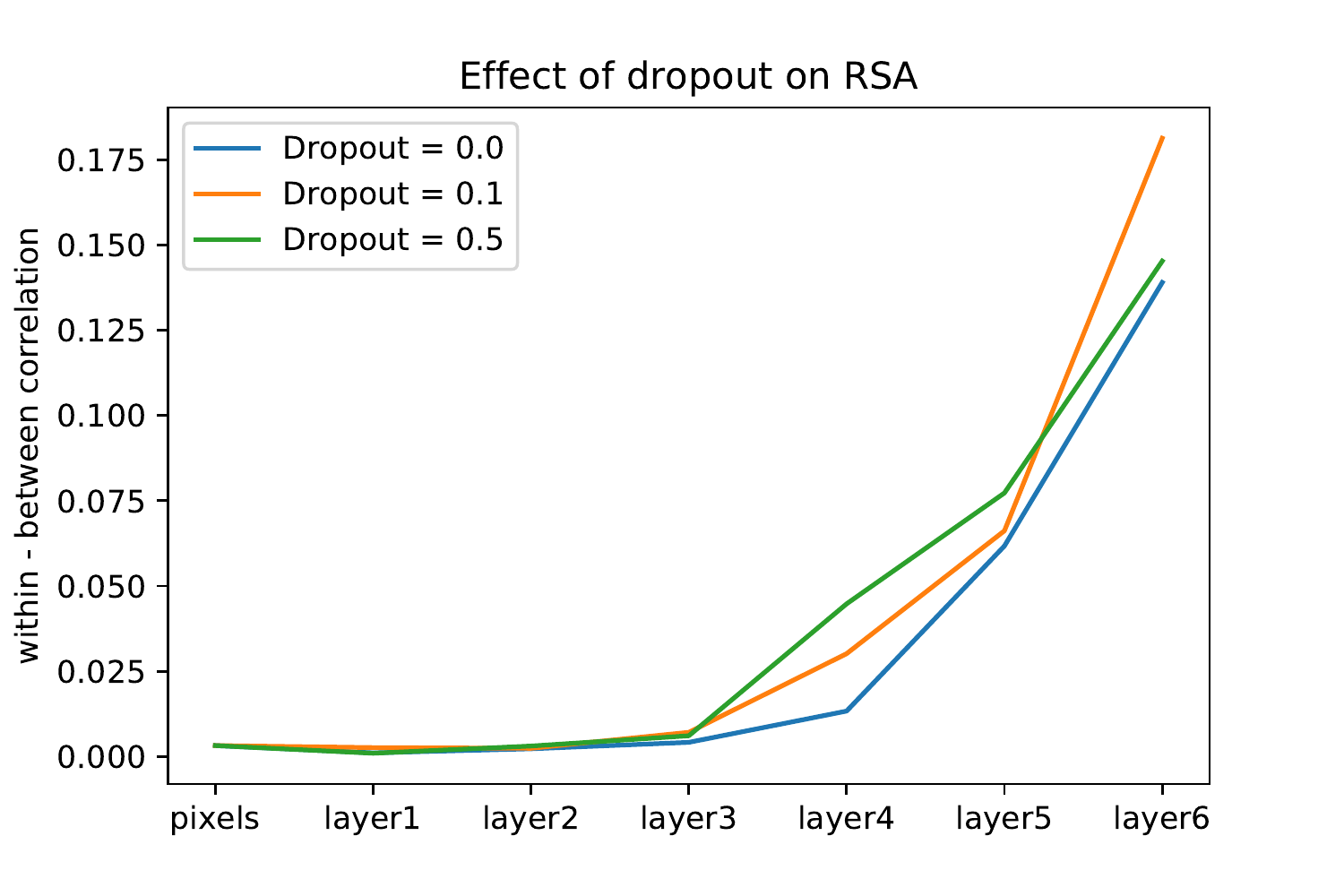}
  \caption{\footnotesize Effect of dropout in PredNet on RSA, as measured by the difference between the average within-category correlation and the average between category correlation (using the Centroid categories derived from human data). Dropout marginally improves the category structure learned in PredNet.}
  \label{fig.prednet_dropout_within_between}
\end{figure}

As discussed in the main paper, our biologically based model includes a number of important biologically motivated properties that may be contributing to the development of its categorical representations. These properties, including excitatory bidirectional connections, inhibitory competition, and an additional form of Hebbian learning, may be acting as regularizers that encourage categorical learning. We therefore tested whether standard regularization methods used in deep learning would have similar effects on the representations developed in the PredNet architecture. We tested 1) batch normalization, 2) dropout (0.1, 0.3, and 0.5), and 3) weight decay (0.01,0.001,0.0001,0.00001). All experiments with batch normalization and weight decay showed reduced performance (in terms of both prediction error on the test set and within-category correlation). As shown in figure \ref{fig.prednet_dropout_within_between}, dropout marginally improved the within-category correlation while also slightly improving prediction accuracy, so a dropout rate of 0.1 was used for the comparison to our biologically based model in the main paper. 

